\documentclass[journal]{IEEEtran}
\usepackage{amsmath}
\usepackage{amssymb}

\usepackage{graphicx}
\usepackage{epstopdf}
\usepackage{amsmath}
\usepackage{array}
\usepackage{tabularx}
\usepackage{multicol}
\usepackage{booktabs}
\usepackage[Symbol]{upgreek}
\usepackage{bm}
\usepackage{cite}
\usepackage{balance}
\usepackage{mathrsfs}
\usepackage{url}
\usepackage{threeparttable}
\usepackage{multirow}
\usepackage{booktabs}
\usepackage{amsfonts}
\usepackage{subfigure}
\usepackage{algorithm}
\usepackage{algorithmic}
\begin{document}
	\title{Dynamic Hybrid Precoding Relying on Twin-Resolution Phase Shifters in Millimeter-Wave Communication Systems}
	\author{\IEEEauthorblockN{Author list}\vspace{-10mm}}
	\author{Chenghao Feng, Wenqian Shen, Xinyu Gao, Jianping An, and Lajos Hanzo,~\IEEEmembership{Fellow,~IEEE}.
        \thanks{This work was supported in part by the National Natural Science Foundation of China (NSFC) under Grant 61620106001, 61901034, and U1836201.
        L. Hanzo would like to acknowledge the financial support of the
Engineering and Physical Sciences Research Council projects
EP/N004558/1, EP/P034284/1, EP/P034284/1, EP/P003990/1 (COALESCE), of
the Royal Society's Global Challenges Research Fund Grant as well as of
the European Research Council's Advanced Fellow Grant QuantCom. (\it{Corresponding author: Wenqian Shen.})}
        \thanks{
		C. Feng, W. Shen, and J. An are with the School of Information and Electronics, Beijing Institute of Technology, Beijing 100081, China (e-mails: cfeng@bit.edu.cn, shenwq@bit.edu.cn, and an@bit.edu.cn). C. Feng is also with Shaanxi Key Laboratory of Information Communication Network and Security, Xi’an University of Posts $\&$ Telecommunications, Xi’an, Shaanxi 710121, China. \par
        X. Gao is with the Huawei Technologies Co. Ltd., Beijing 100085, China (e-mail: gaoxinyu@huawei.com).\par
        L. Hanzo is with the Department of Electronics and Computer Science, University of Southampton, Southampton SO17 1BJ, UK (e-mail: lh@ecs.soton.ac.uk).
		}
	}
	\maketitle
	\begin{abstract}
    Hybrid analog/digital precoding in millimeter-wave (mmWave) multi-input multi-ouput (MIMO) systems is capable of achieving the near-optimal full-digital performance at reduced hardware cost and power consumption compared to its full-RF digital counterpart.
    However, having numerous phase shifters is still costly, especially when the phase shifters are of high resolution.
    In this paper, we propose a novel twin-resolution phase-shifter network for mmWave MIMO systems, which reduces the power consumption of an entirely high-resolution network, whilst mitigating the severe array gain reduction of an entirely low-resolution network.
    The connections between the twin phase shifters having different resolutions and the antennas are either fixed or dynamically configured.
    In the latter, we jointly design the phase-shifter network and the hybrid precoding matrix, where the phase of each entry in the analog precoding matrix can be dynamically designed according to the required resolution.
    This method is slightly modified for the fixed network's hybrid precoding matrix.
    Furthermore, we extend the proposed method to multi-user MIMO systems and provide its performance analysis.
    Our simulation results show that the proposed dynamic hybrid precoding method strikes an attractive performance vs. power consumption trade-off.
    %Our simulation results show that the proposed dynamic hybrid precoding method strikes an attractive trade-off between the power consumption and system performance.
	\end{abstract}
	\IEEEpeerreviewmaketitle
	\section{Introduction}\label{S1}
    \IEEEPARstart{M}{illimeter}-wave (mmWave) solutions have become one of the key techniques for next-generation wireless communication systems \cite{8337813,Shen,JTSP_HRobert_OverviewMmwave}.
    Having plenty of bandwidth at mmWave frequencies relieves the scarcity of spectral resources caused by the explosive growth of data traffic and electronic devices \cite{8198807}.
    Nevertheless, mmWave signals inevitably suffer from high path loss \cite{7109864}.
    Fortunately, the short wavelength of mmWave carriers enables the application of compact antennas and further guarantees the integration of large-scale antenna arrays in a compact physical size \cite{8463485,8777168,8698333,6894454,5262293,8370683}.
    Consequently, massive multi-input multi-output (MIMO) schemes constitute promising techniques for mmWave communication systems, where the high beamforming gain of large antenna arrays is capable of counteracting the high path loss \cite{6979962,7010533,6736750}.
    However, in conventional fully digital MIMO structures, each antenna is supported by a dedicated radio-frequency (RF) chain, which will lead to unaffordable hardware cost and power consumption.
    \subsection{Literature review}\label{S1.1}
    Hybrid precoding comes to rescue in circumventing this problem \cite{el2013spatially,7579557,8457260,Alkhateeb2016Frequency,TWC_AAhmed_LimitedHybridPrecoding,7742917}.
    It divides the fully digital precoding into two parts, including analog precoding and digital precoding.
    The transmit signals are processed by precoding both in the digital and analog domains for attaining both high multiplexing and beamforming gains.
    In \cite{el2013spatially,7579557,8457260,Alkhateeb2016Frequency}, hybrid precoding techniques have been conceived for mmWave point-to-point MIMO systems.
    El-Ayach \textit{et al}. in \cite{el2013spatially} propose an orthogonal matching pursuit (OMP)-based method to design the hybrid precoder.
    Rusu \textit{et al}. in \cite{7579557} propose an optimization method for designing hybrid precoders by minimizing the Euclidean distance between the fully digital solution and the hybrid precoder.
    Zhang \textit{et al}. in \cite{8457260} adopt penalty decomposition methods for designing the hybrid precoder of wideband systems relying on orthogonal frequency division multiplexing (OFDM).
    Alkhateeb \textit{et al}. in \cite{Alkhateeb2016Frequency} propose efficient hybrid analog/digital codebooks and a corresponding near-optimal hybrid precoding method for wideband mmWave systems.
    Furthermore, the authors of \cite{TWC_AAhmed_LimitedHybridPrecoding,7742917} study the benefit of hybrid precoding methods in mmWave multi-user MIMO (MU-MIMO) systems.
    A codebook based method is proposed in \cite{TWC_AAhmed_LimitedHybridPrecoding}, where the columns of the analog precoding matrix are selected from a codebook for maximizing the bandwidth efficiency, while the digital precoder is designed based on the zero-forcing (ZF) method for eliminating the interference among users.
    %The authors in \cite{7959183} design the hybrid precoding on basis of the concave-convex optimization and weighted mean-square error minimization in MU-MIMO systems.
    The authors of \cite{7742917} design a hybrid precoder for a scenario, when each user is supported by several RF chains.
    An iterative matrix decomposition-aided block diagonalization method is proposed for eliminating the inter-user interference.
   % A hybrid precoding method based on matrix-monotonic optimization is proposed in \cite{8695847} to jointly design the analog and digital precoding.

    However, the aforementioned hybrid precoding methods assume that phase shifters are of infinite or high resolutions.
    Since numerous phase shifters are required, they impose high power consumption and hardware complexity.
    Therefore, using low-resolution, low-cost phase shifters is preferred for practical systems \cite{8303761,8332507,7858800,7227015,7389996,8422249,8331836}.
%    The authors in \cite{7858800} propose an iterative algorithm to design the phases of the analog or RF beamformers for spectral efficiency maximization
    Chen \textit{et al}. in \cite{8303761} propose a hybrid transmit precoding (TPC) scheme based on joint iterative training and low-resolution phase shifters, where the iterative training used converges to the dominant steering vectors that align with the direction of the highest channel gain.
    Shi and Hong in \cite{8332507} propose a penalty dual decomposition method for designing hybrid TPC aiming for maximizing the bandwidth efficiency.
    Chen in \cite{7858800} proposes an iterative algorithm for finding the desired discrete phases that maximize the bandwidth efficiency.
    Sohrabi and Yu in \cite{7227015,7389996} conceive a heuristic hybrid beamforming design, which achieves near-optimal performance in terms of throughput in point-to-point MIMO systems and in multi-user multiple-input single-output (MU-MISO) systems.
    Wang \textit{et al}. in \cite{8422249,8331836} propose an iterative algorithm, which successively designs the low-resolution analog precoder and combiner, aiming for conditionally maximizing the bandwidth efficiency.
    %These works design appropriate digital precoder to suppress the inter-user interference.
    Nevertheless, all phase shifters are considered to be of low resolution, which leads to severe array-gain loss.
    In addition to using phase shifters that are digitally controlled, passive analog precoders are adopted for constructing hybrid precoding.
    Tan \textit{et al.} \cite{7956161} adopt a discrete Fourier transform (DFT)-based analog precoder and investigate the achievable rate of MU-MISO systems in Rayleigh fading channels.
    Furthermore, Han \textit{et al.} \cite{8327819} utilize the cost-effective Butler matrices for enhancing DFT-based systems and propose a two-step hybrid precoding scheme for MU-MIMO systems, which achieve near-optimal performances despite substantially reducing the complexity of the conventional exhaustive search.
    However, compared with phase shifter based systems, the bandwidth efficiency of passive analog precoders is limited when a small amount of RF chains are adopted.

    \subsection{Contributions}\label{S1.2}
    We propose a twin-resolution phase-shifter network together with a dynamic hybrid TPC method for striking a compelling performance vs. power consumption trade-off.
    The main contributions of this paper are summarized as follows:
    \begin{enumerate}
    \item
    For simplicity, we assume that the numbers of high- and low-resolution phase shifters are equal to each other.
    Thus, we propose to replace half of the high-resolution phase shifters with low-resolution ones in the conventional entirely high-resolution network and acquire our proposed twin-resolution network, which reduces the power consumption of the entirely high-resolution network and compensates for the severe array gain loss of the entirely low-resolution network.
    Two types of twin-resolution phase-shifter networks are considered, namely a fixed network and a dynamic network.
    For the fixed network, we fix the connections between the twin-resolution phase shifters and transmit antennas (TAs), while in the dynamic network, the connections are dynamically configured with the aid of a mapper between the twin-resolution phase shifters and TAs.
    Note that a certain TA can be flexibly connected to either a high- or low-resolution phase shifter in the dynamic network.
    \item
    For the proposed dynamic twin-resolution phase-shifter networks, we aim for maximizing the bandwidth efficiency and propose a dynamic hybrid TPC method by jointly designing the mapper in the network and the hybrid TPC matrix.
    Specifically, we derive the relationship between the bandwidth efficiency and each entry in the analog TPC matrix.
    Then we propose an iterative method for jointly designing the entries in the analog TPC matrix and the mapper.
    Specifically, we first determine the mapper for low-resolution phase shifters and quantize the phases of the corresponding entries in the analog TPC at a low resolution.
    Afterwards, we design the phases of the remaining entries having high resolution.
    The proposed dynamic hybrid TPC method can be slightly modified for employment, when a fixed phase-shifter network with fixed mapper is adopted.
    \item
    In MU-MIMO systems, we eliminate the inter-user interference by the classic block diagonalization method in the baseband TPC matrix design.
    We further characterize the bandwidth efficiency in a form, which is composed of the bandwidth efficiency in the absence of inter-user interference and the bandwidth efficiency erosion caused by inter-user interference.
    We propose to jointly design the analog TPC matrix and the mapper by maximizing the former.
    Our simulation results demonstrate that the proposed dynamic hybrid TPC method achieves higher energy efficiency (EE) than its traditional counterparts.
    \item
    Furthermore, we provide the performance analysis of the proposed method. To derive the bandwidth efficiency gap between the proposed method relying on our twin-resolution phase-shifter network and the fully digital solution, we sequentially replace the entries of the fully digital TPC matrix by that of the hybrid TPC matrix derived.
    The bandwidth efficiency gap equals to the summation of all bandwidth efficiency variations during the process of replacement.
    \end{enumerate}
    The remainder of this paper is organized as follows.
    In Section \ref{S2}, our system and channel models are introduced.
    In Section \ref{S3}, our twin-resolution phase-shifter networks are presented.
    In Section \ref{S4}, we propose our dynamic hybrid TPC method for mmWave point-to-point MIMO systems and provide its performance analysis.
    In Section \ref{S5}, we extend the proposed method and performance analysis to MU-MIMO systems.
    In Section \ref{S6}, numerical results are provided.
    Finally, our conclusions are drawn in Section \ref{S7}.

    \emph{Notation}:
	Lower-case and upper-case boldface letters denote vectors and matrices, respectively.
	$(\cdot)^{\rm{T}}$, $(\cdot)^{\text{H}}$, $(\cdot)^{-1}$ and $(\cdot)^{\dagger}$ denote the transpose, conjugate transpose, inverse and pseudo-inverse of a matrix, respectively.
    $tr(\cdot)$ presents the trace function.
    $\mathrm{diag}(\cdot)$ extracts the diagonal element of a matrix into a column vector.
    $\| \cdot \|_{F}$ denotes the Frobenius norm of a matrix.
    $|a|$ is the absolute value of a scalar.
    $|\mathbf{A}|$ is the determinant of a matrix.
    $\mathbf{A}_{\left(i,:\right)}$ and $\mathbf{A}_{\left(:,j\right)}$ represent the $i$-th row and $j$-th column of the matrix $\mathbf{A}$, respectively.
    The operator $\circ$ denotes the Hadamard product.
	Finally, $\mathbf{I}_P$ denotes the identity matrix of size $P\times P$.
	\section{Downlink System and Channel Model}\label{S2}
    In this section, we will introduce our system model and channel model for both mmWave point-to-point MIMO and MU-MIMO systems.
	\subsection{MmWave Point-to-point MIMO Systems}\label{S2.1}
\begin{figure}[t]
		\center{\includegraphics[width=0.45\textwidth]{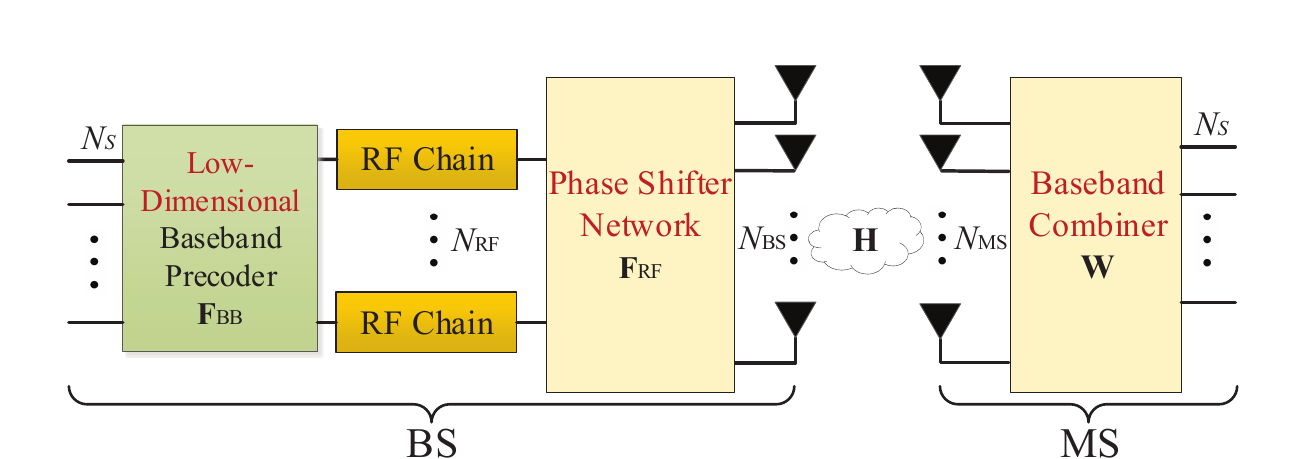}}
		\caption{Illustration of mmWave point-to-point MIMO systems.}
		\label{transceiver_SU}
	\end{figure}

Firstly, we will introduce the mmWave point-to-point MIMO downlink (DL) considered.
As shown in Fig. \ref{transceiver_SU}, the base station (BS) communicates with the user through $N_\mathrm{BS}$ TAs and $N_\mathrm{RF}$ RF chains.
The user is equipped with $N_\mathrm{MS}$ receiver antennas (RAs) (fully digital structure).
The number of data streams is $N_s$.
The transmit signal is defined as $\mathbf{\mathbf{s}} = \left[ s_1,s_2,\cdots,s_{N_s} \right]^{\text{T}} \in \mathbb{C}^{N_s \times 1}$ with normalized power, i.e., $\mathbb{E}\left[{\mathbf{s}\mathbf{s}^{\text{H}}}\right] = \mathbf{I}_{N_s}$.
The transmit signal at the BS is firstly precoded by the low-dimensional baseband TPC $\mathbf{F}_{\mathrm{BB}} \in \mathbb{C}^{N_{\mathrm{RF}} \times N_s}$.
Then, the digitally-precoded signals $\mathbf{F}_{\mathrm{BB}}\mathbf{s}$ are processed by the analog precoder, which is accomplished by the phase-shifter network.
There are $N_\mathrm{RF}$ groups of phase shifters and each group has $N_\mathrm{BS}$ phase shifters.
The phase-shifter network will be discussed in Section \ref{S3}.
The analog TPC is expressed in a matrix form, i.e., $\mathbf{F}_{\mathrm{RF}} \in \mathbb{C}^{N_\mathrm{BS} \times N_{\mathrm{RF}}}$.
Due to the hardware constraint of phase shifters, the entries of the analog TPC matrix are discrete as in $\mathbf{F}_{\mathrm{RF}}\left(i,j \right) = \frac{1}{\sqrt{N_\mathrm{BS}}} e^{j\phi_{i,j}}$, where $\phi_{i,j}$ is the discrete phase caused by the limited resolution of the phase shifters.
The hybrid TPC satisfies the power constraint of $\| \mathbf{F}_{\mathrm{RF}}\mathbf{F}_{\mathrm{BB}} \|_F^2 = N_s$.
After hybrid TPC, the transmit signal is expressed as
     \begin{align}\label{s_BS}
    \mathbf{x} = \sqrt{\frac{\rho}{N_s}} \mathbf{F}_{\mathrm{RF}}\mathbf{F}_{\mathrm{BB}}\mathbf{s},
	\end{align}
where $\rho$ is the transmit power.

A narrow-band mmWave channel model $\mathbf{H} \in \mathbb{C}^{N_{\mathrm{MS}} \times N_{\mathrm{BS}}}$ is adopted where uniform planar arrays (UPAs) of antennas are employed both at the BS and the user.
We adopt the physical multi-path channel model having $L$ paths for narrow-band mmWave systems, which is widely used in \cite{el2013spatially,7389996,8331836},
    \begin{align}\label{H}
    \mathbf{H} = \sqrt{\frac{N_\mathrm{BS}N_\mathrm{MS}}{L}} \sum_{\ell = 1}^{L}\alpha_{\ell}\mathbf{a}_{\mathrm{MS}}\left( \phi^{\mathrm{r}}_{\ell}, \theta^{\mathrm{r}}_{\ell} \right)\mathbf{a}_{\mathrm{BS}}^{\text{H}}\left( \phi^{\mathrm{t}}_{\ell}, \theta^{\mathrm{t}}_{\ell} \right),
	\end{align}
where $\alpha_{\ell}\sim \mathcal{CN}\left(0, 1\right)$ is the complex gain of the $\ell$-th path.
The vectors $\mathbf{a}_{\mathrm{BS}}\left( \phi^{\mathrm{t}}_{\ell}, \theta^{\mathrm{t}}_{\ell} \right) \in \mathbb{C}^{N_{\mathrm{BS}} \times 1 }$ and $\mathbf{a}_{\mathrm{MS}}\left( \phi^{\mathrm{r}}_{\ell}, \theta^{\mathrm{r}}_{\ell} \right)\in \mathbb{C}^{N_{\mathrm{MS}} \times 1 }$ represent the antenna array responses at the BS and the user, which can be expressed as
    \begin{align}\label{a_BS}
    \mathbf{a}_{\mathrm{BS}}\left( \phi^{\mathrm{t}}_{\ell}, \theta^{\mathrm{t}}_{\ell} \right) = & \frac{1}{\sqrt{N_{\mathrm{BS}}}}  \left[ 1,\cdots, e^{j\frac{2\pi}{\lambda}d\left( m \mathrm{sin} \phi^{\mathrm{t}}_{\ell} \mathrm{sin} \theta^{\mathrm{t}}_{\ell} + n \mathrm{cos}\theta^{\mathrm{t}}_{\ell} \right) } ,\cdots, \right. \nonumber\\
    &\quad \quad \left. e^{j\frac{2\pi}{\lambda}d\left( \left( W-1 \right) \mathrm{sin} \phi^{\mathrm{t}}_{\ell} \mathrm{sin} \theta^{\mathrm{t}}_{\ell} + \left( H-1 \right) \mathrm{cos}\theta^{\mathrm{t}}_{\ell} \right) } \right]^{\text{T}},
	\end{align}
where $\phi^{\mathrm{t}}_{\ell}$ and $\theta^{\mathrm{t}}_{\ell}$ denote the azimuth and elevation angle-of-departure (AoD), $\lambda$ represents the wavelength, and $d$ is the antenna spacing.
Furthermore, $m \in \{0,1,\cdots,W-1 \}$ and $n \in \{0,1,\cdots,H-1 \}$ with $W$ and $H$ denoting the number of antennas in horizontal and vertical directions.
Similarly, $\mathbf{a}_{\mathrm{MS}}\left( \phi^{\mathrm{r}}_{\ell}, \theta^{\mathrm{r}}_{\ell} \right)$ can be expressed in the same form as (\ref{a_BS}) by replacing $\phi^{\mathrm{t}}_{\ell}$ and $\theta^{\mathrm{t}}_{\ell}$ with $\phi^{\mathrm{r}}_{\ell}$ and $\theta^{\mathrm{r}}_{\ell}$.
Typically, we set $N_\mathrm{s} = \min\left\{ L, N_\mathrm{RF} \right\}$.

The received DL signal is expressed as
    \begin{align}\label{y}
    \mathbf{y} =  \sqrt{\frac{\rho}{N_s}} \mathbf{H}\mathbf{F}_{\mathrm{RF}}\mathbf{F}_{\mathrm{BB}}\mathbf{s} + \mathbf{n} ,
	\end{align}
    where $ \mathbf{n} \sim \mathcal{CN}\left(0, \sigma^2\mathbf{I} \right) \in \mathbb{C}^{N_{\mathrm{MS}} \times 1}$.
The received signal at the user is combined by the full-digital combiner $\mathbf{W} \in \mathbb{C}^{N_{\mathrm{MS}} \times N_s}$.
Thus, we have
    \begin{align}\label{r}
    \mathbf{r} = \sqrt{\frac{\rho}{N_s}}\mathbf{W}^{\text{H}}\mathbf{H}\mathbf{F}_{\mathrm{RF}}\mathbf{F}_{\mathrm{BB}}\mathbf{s} + \mathbf{W}^{\text{H}}\mathbf{n} .
	\end{align}
The bandwidth efficiency $\mathbf{\emph{R}}$ achieved with Gaussian signaling over the mmWave point-to-point MIMO channel can be expressed as \cite{el2013spatially}
    \begin{align}\label{R}
     \mathbf{\emph{R}} = \mathrm{log}_2  \left(\left| \mathbf{I}_{N_s} + \frac{\rho}{N_s\sigma^2}\mathbf{R}^{-1} \mathbf{W}^{\text{H}} \mathbf{H} \mathbf{F}_{\mathrm{RF}} \mathbf{F}_{\mathrm{BB}}  \mathbf{F}^{\text{H}}_{\mathrm{BB}} \mathbf{F}^{\text{H}}_{\mathrm{RF}} \mathbf{H}^{\text{H}} \mathbf{W} \right|\right) ,
	\end{align}
    where $\mathbf{R} = \mathbf{W}^{\text{H}}\mathbf{W}$ is the noise covariance matrix after combining.
    \subsection{MmWave MU-MIMO Systems}\label{S2.2}
 	\begin{figure}[t]
		\center{\includegraphics[width=0.45\textwidth]{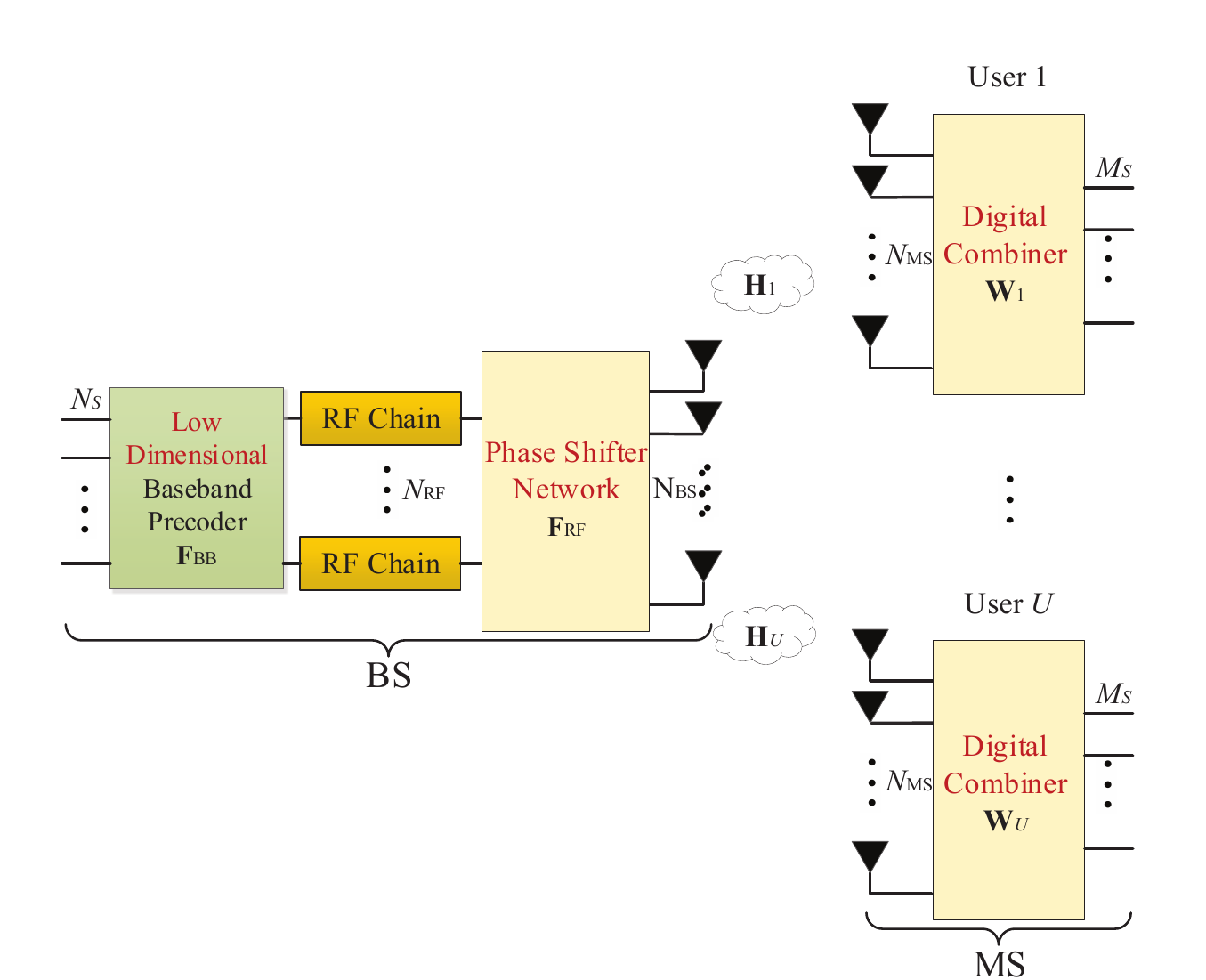}}
		\caption{Illustration of mmWave MU-MIMO systems.}
		\label{transceiver_MU}
	\end{figure}
    In this subsection, we will consider mmWave MU-MIMO DL systems.
    As shown in Fig. \ref{transceiver_MU}, the BS transmits in the DL to $U$ users through $N_\mathrm{BS}$ TAs and $N_\mathrm{RF}$ RF chains.
    Each user is equipped with $N_\mathrm{MS}$ RAs (fully digital structure).
    The number of data streams at the BS is $N_s = UM_s$, where $M_s$ is the number of data streams at each user.
    In our MU-MIMO systems, the DL transmit signal is $\mathbf{\mathbf{s}} = \left[ s_1,\cdots,s_{M_s} ,\cdots, s_{U\left(M_s-1\right)+1},\cdots, s_{UM_s} \right]^{\text{T}} \in \mathbb{C}^{N_s \times 1}$.
    The baseband TPC at the BS is divided into $U$ submatrices and we have $\mathbf{F}_{\mathrm{BB}} = \left[ \mathbf{F}^1_{\mathrm{BB}},\mathbf{F}^2_{\mathrm{BB}},\cdots,\mathbf{F}^{U}_{\mathrm{BB}} \right] \in \mathbb{C}^{N_s \times UM_s}$, where each submatrix is defined as $\mathbf{F}^u_{\mathrm{BB}} = \mathbf{F}_{\mathrm{BB}\left(:,\left( u-1 \right)M_s+1:uM_s\right)}$.
    The hybrid TPC satisfies the power constraint of $\| \mathbf{F}_{\mathrm{RF}}\mathbf{F}_{\mathrm{BB}} \|_F^2 = N_s$.

    The DL channel $\mathbf{H}_u \in \mathbb{C}^{N_{\mathrm{MS}} \times N_{\mathrm{BS}}}$ between the BS transmitter and the $u$-th receiver is expressed as
    \begin{align}\label{H_u}
    \mathbf{H}_u = \sqrt{\frac{N_\mathrm{BS}N_\mathrm{MS}}{L_u}} \sum_{\ell = 1}^{L_u}\alpha_{u,\ell}\mathbf{a}_{\mathrm{MS}}\left( \phi^{\mathrm{r}}_{u,\ell}, \theta^{\mathrm{r}}_{u,\ell} \right)\mathbf{a}_{\mathrm{BS}}^{\text{H}}\left( \phi^{\mathrm{t}}_{u,\ell}, \theta^{\mathrm{t}}_{u,\ell} \right),
	\end{align}
    where $L_u$ and $\alpha_{u,\ell}$ are the number of propagation paths and complex path gains.
    The vectors $\mathbf{a}_{\mathrm{MS}}\left( \phi^{\mathrm{r}}_{u,\ell}, \theta^{\mathrm{r}}_{u,\ell} \right)$ and $\mathbf{a}_{\mathrm{BS}}\left( \phi^{\mathrm{t}}_{u,\ell}, \theta^{\mathrm{t}}_{u,\ell} \right)$ represent the antenna array responses at the $u$-th user and the BS.

    At the receiver, each user combines the received signal with the aid of the full-digital combiner $\mathbf{W}_u \in \mathbb{C}^{N_{\mathrm{MS}} \times M_s}$.
    The received signal at the $u$-th user after combining is expressed as
    \begin{align}\label{r_u}
     \mathbf{r}_u = \sqrt{\frac{\rho}{N_s}}\mathbf{W}_u^{\text{H}}\mathbf{H}_u\mathbf{F}_{\mathrm{RF}}\mathbf{F}_{\mathrm{BB}}\mathbf{s} + \mathbf{W}_u^{\text{H}}\mathbf{n}_u .
    \end{align}
    The bandwidth efficiency $\mathbf{\emph{R}}_u$ for the $u$-th user achieved by Gaussian signaling over the mmWave MU-MIMO channel is expressed as \cite{8303761}
    \begin{align}\label{R_u}
    \mathbf{\emph{R}} = & \sum_{u=1}^U \mathbf{\emph{R}}_u \nonumber \\
    = & \sum_{u=1}^U \mathrm{log}_2 \left( \left| \mathbf{I}_{M_s} + \frac{\rho}{N_s}\mathbf{R}_i^{-1} \mathbf{W}_{u}^{\text{H}} \mathbf{H}_u \mathbf{F}_{\mathrm{RF}} \mathbf{F}^u_{\mathrm{BB}} \right.\right.\nonumber\\
    & \quad \quad \quad \quad \quad \left.\left. \left(\mathbf{F}^u_{\mathrm{BB}}\right)^{\text{H}} \mathbf{F}^{\text{H}}_{\mathrm{RF}} \mathbf{H}_u^{\text{H}} \mathbf{W}_{u} \right| \right),
	\end{align}
    where $\mathbf{R}_i = \sum_{i=1,i \neq u}^U\mathbf{W}_{u}^{\text{H}} \mathbf{H}_u \mathbf{F}_{\mathrm{RF}} \mathbf{F}^i_{\mathrm{BB}} \left(\mathbf{F}^i_{\mathrm{BB}}\right)^{\text{H}} \mathbf{F}^{\text{H}}_{\mathrm{RF}} \mathbf{H}_u^{\text{H}}$ $\mathbf{W}_{u} + \sigma^2 \mathbf{W}_{u}^{\text{H}}\mathbf{W}_{u}$ represents the interference-plus-noise covariance matrix.
    \section{Proposed Twin-Resolution Phase-Shifter Network}\label{S3}
    In this subsection, we will propose two types of twin-resolution phase-shifter networks.
    \subsection{Fixed Twin-Resolution Phase-Shifter Network}\label{S3.1}
	\begin{figure}[t]
		\center{\includegraphics[width=0.45\textwidth]{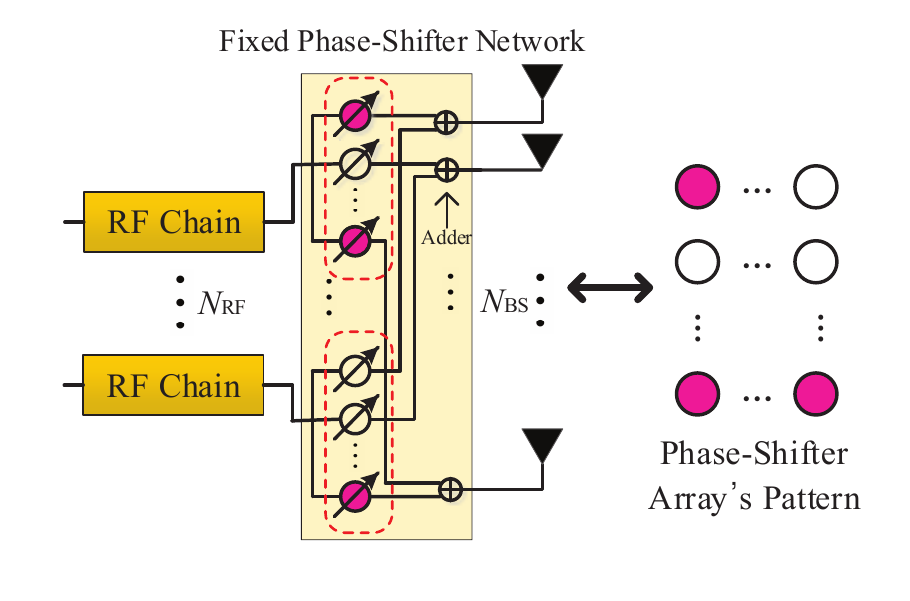}}
		\caption{Illustration of the proposed fixed twin-resolution phase-shifter network and phase-shifter array's pattern.}
		\label{fixed twin-resolution}
	\end{figure}
    \begin{figure}[t]
    \centering
    \subfigure[ ]{
    \begin{minipage}[t]{0.1\textwidth}
    \includegraphics[width=1\textwidth]{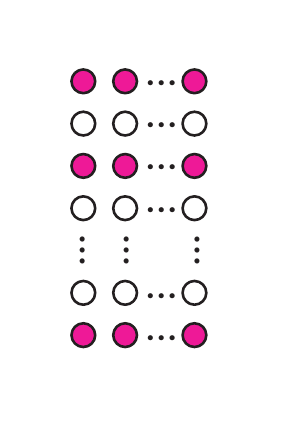}
    \end{minipage}
    }
    \subfigure[ ]{
    \begin{minipage}[t]{0.1\textwidth}
    \includegraphics[width=1\textwidth]{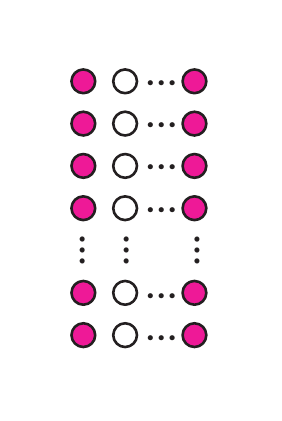}
    \end{minipage}
    }
    \subfigure[ ]{
    \begin{minipage}[t]{0.1\textwidth}
    \includegraphics[width=1\textwidth]{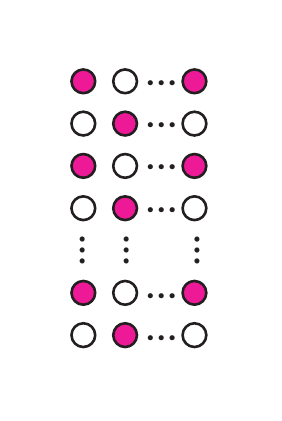}
    \end{minipage}
    }
    \subfigure[ ]{
    \begin{minipage}[t]{0.1\textwidth}
    \includegraphics[width=1\textwidth]{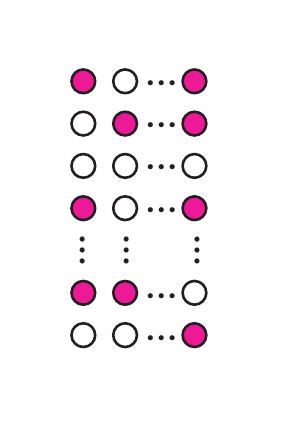}
    \end{minipage}
    }
    \caption{Illustration of the fixed twin-resolution phase-shifter array's pattern: (a) horizontal; (b) vertical; (c) interlaced; (d) random.}
    \label{structure_d}
    \end{figure}
    In order to reduce the hardware cost and power consumption, while maintaining high bandwidth efficiency, we propose to utilize twin-resolution phase shifters.
    For simplicity, both the number of high- and low-resolution phase shifters are assumed to be half the total.
    We use the phase-shifter array pattern of Fig. \ref{fixed twin-resolution} for characterizing the phase shifter network.
    As shown in Fig. \ref{fixed twin-resolution}, the colored circles in the pattern correspond to the colored high-resolution phase shifters, while the hollow circles represent the low-resolution phase shifters.
    The phase-shifter array pattern is of size $N_{\mathrm{BS}}\times N_s$.
    The $j$-th column in the pattern represents the corresponding phase shifter group connected to the $j$-th RF chain.
    In each column, the circle at the $i$-th row corresponds to the phase shifter connected to the $i$-th antenna.
    Furthermore, in Fig. \ref{structure_d}, we show four typical twin-resolution phase-shifter array patterns.
    Fig. \ref{structure_d} (a) shows the horizontal pattern, where all phase shifter groups have the same pattern.
    Fig. \ref{structure_d} (b) shows a vertical pattern.
    In this pattern, the phase shifter group connected to a certain RF chain is either entirely high-resolution or entirely low-resolution.
    Furthermore, an interlaced pattern is shown in Fig. \ref{structure_d} (c), where any two adjacent points have different resolutions.
    The above three phase-shifter array patterns are regular.
    By contrast, a randomly generalized pattern is shown in Fig. \ref{structure_d} (d).
    Next, we will propose a dynamic twin-resolution phase-shifter network in the next subsection, where the phase-shifter array pattern can be dynamically reconfigured.
    \subsection{Dynamic Twin-Resolution Phase-Shifter Network}\label{S3.2}
 	\begin{figure}[t]
		\center{\includegraphics[width=0.45\textwidth]{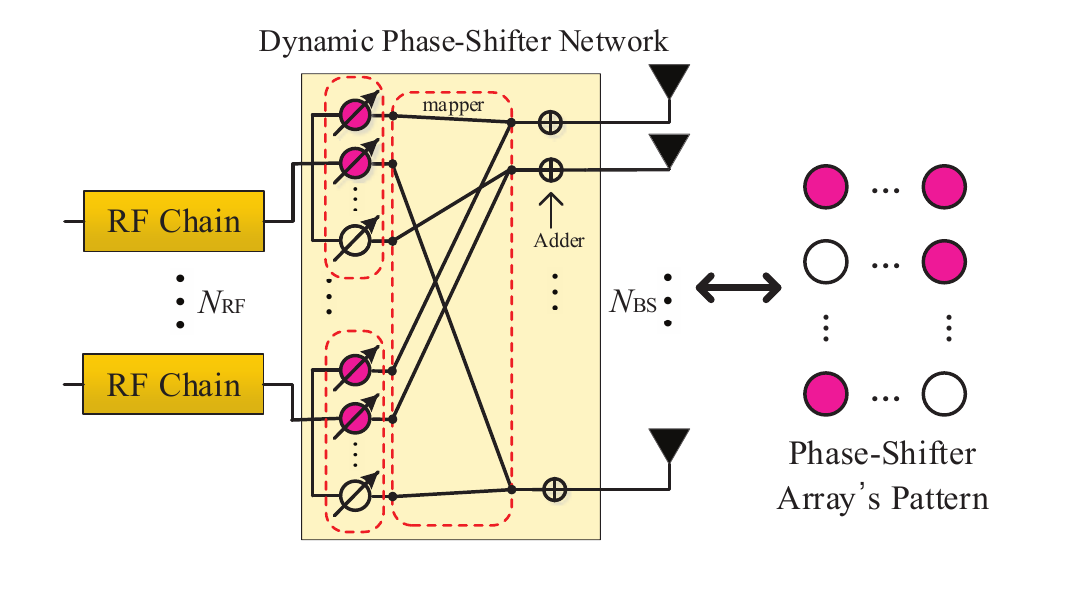}}
		\caption{Illustration of the proposed dynamic twin-resolution phase-shifter network.}
		\label{dynamic twin-resolution}
	\end{figure}
    In this subsection, we propose a dynamic phase-shifter network.
    As illustrated in Fig. \ref{dynamic twin-resolution}, each phase shifter group contains $\frac{N_{\mathrm{BS}}}{2}$ high-resolution phase shifters and $\frac{N_{\mathrm{BS}}}{2}$ low-resolution phase shifters.
    Distinguished from the fixed network, the mapper in our dynamic network can be dynamically configured, which can be characterized by the phase-shifter array pattern as well.
    For instance, the corresponding pattern is depicted in Fig. \ref{dynamic twin-resolution}.

    The dynamic mapper can be implemented by providing switches between the phase shifters and the TAs, which is motivated by \cite{7370753,7394147,7446253}.
    In our dynamic twin-resolution phase-shifter network, each phase shifter is connected to $N_{\mathrm{BS}}$ ``on-off" switches, which can connect the phase shifter to the required TA.
The total number of required switches is $N_{\mathrm{BS}}^2N_s$.
However, during the transmission, only one of the $N_{\mathrm{BS}}$ switches is activated for connecting a phase shifter to a TA, as illustrated in Fig. 5.
The power consumption of switches, which remain ``off", is negligible compared to that of the active switches \cite{8903536}.
Therefore, the number of active switches is $N_{\mathrm{BS}}N_s$.
Moreover, some recent studies on switch-based systems justify the feasibility of implementing large number of switches in MIMO systems \cite{8903536,9026753,7880698}.
    The implementation of switches has low power consumption and low insertion loss \cite{7880698,7387790,8295113,8310586}.
    Furthermore, the dynamic twin-resolution phase-shifter network strikes a beneficial performance vs. complexity trade-off  \cite{8310586}, the high bandwidth efficiency and energy efficiency brought about by the dynamic phase-shifter network indicates that the switches are indeed beneficial as detailed in Section \ref{S6}.

    In the twin-resolution phase-shifter networks, the number of quantization bits in the high-resolution phase shifters is denoted by $B_{\mathrm{H}}$ and that of the low-resolution phase shifters is denoted by $B_{\mathrm{L}}$.
    The discrete phases of the high-resolution and low-resolution phase shifters are included in the sets $\mathcal{Q}_{\mathrm{H}}$ and $\mathcal{Q}_{\mathrm{L}}$, respectively.
    Furthermore, we define the index sets of high-resolution and low-resolution phase shifters in the phase-shifter array pattern as $\mathcal{S}_{\mathrm{H}}$ and $\mathcal{S}_{\mathrm{L}}$, respectively.
    We have $\left| \mathcal{S}_{\mathrm{H}}\right| + \left|\mathcal{S}_{\mathrm{L}} \right| = N_{\mathrm{BS}}N_s$.

    In the next section, we will propose a dynamic hybrid TPC method for the dynamic phase-shifter network, which can be slightly modified to construct a fixed network.
    \section{Proposed Dynamic Hybrid Precoding for MmWave Point-to-point MIMO Systems}\label{S4}
    We make the idealized simplifying assumption that the downlink CSI is known \cite{7880698}.
    In order to decouple the TPC and combiner design at the BS and the user, we assume that the user employs the optimal unconstrained combiner $\mathbf{W}_{\mathrm{opt}}=\mathbf{U}_{\left(:, 1:N_s \right)}$ \cite{7579557}, where $\mathbf{U}$ is the left singular matrix of the channel matrix $\mathbf{H}$.
    Consequently, the bandwidth efficiency in (\ref{R}) is rewritten as
    \begin{align}\label{R_opt}
    \mathbf{\emph{R}} = \mathrm{log}_2 \left( \left| \mathbf{I}_{N_s} + \frac{\rho}{N_s\sigma^2} \mathbf{W}_{\mathrm{opt}}^{\text{H}} \mathbf{H} \mathbf{F}_{\mathrm{RF}} \mathbf{F}_{\mathrm{BB}} \mathbf{F}^{\text{H}}_{\mathrm{BB}} \mathbf{F}^{\text{H}}_{\mathrm{RF}} \mathbf{H}^{\text{H}} \mathbf{W}_{\mathrm{opt}} \right|\right) .
	\end{align}
    The hybrid TPC design is divided into two steps.
    Firstly, for a given analog TPC $\mathbf{F}_{\mathrm{RF}}$, the digital TPC is designed for eliminating the interference among data streams. Afterwards, we iteratively design the entries of $\mathbf{F}_{\mathrm{RF}}$, which is realized by the twin-resolution phase shifters.
    We formulate the hybrid TPC design problem as
    %\begin{align}
    \begin{subequations}\label{ObjSU}
    \begin{align}
     \max_{\mathbf{F}_{\mathrm{RF}},\mathbf{F}_{\mathrm{BB}}} \mathrm{log}_2 \left( \left| \mathbf{I}_{N_s} + \frac{\rho}{N_s\sigma^2} \mathbf{W}_{\mathrm{opt}}^{\text{H}} \mathbf{H} \mathbf{F}_{\mathrm{RF}} \mathbf{F}_{\mathrm{BB}} \right.\right.\nonumber\\ \left.\left. \mathbf{F}^{\text{H}}_{\mathrm{BB}} \mathbf{F}^{\text{H}}_{\mathrm{RF}} \mathbf{H}^{\text{H}} \mathbf{W}_{\mathrm{opt}} \right|\right)
    \end{align}
    \begin{equation}
    s.t. \quad \| \mathbf{F}_{\mathrm{RF}}\mathbf{F}_{\mathrm{BB}} \|_F^2 = N_s,
    \end{equation}
    \begin{equation}
    \quad \quad \quad \left|\mathbf{F}_{\mathrm{RF}}\left(i,j \right)\right| = \frac{1}{\sqrt{N_\mathrm{BS}}}.
    \end{equation}
    \end{subequations}
    For a given $\mathbf{F_{\rm RF}}$, we adopt the baseband TPC matrix $\mathbf{F}_{\rm BB}$ according to the popular SVD-based method \cite{el2013spatially} and multiply it by a coefficient as
     \begin{align}\label{FBB}
    \mathbf{F}_{\mathrm{BB}} = \sqrt{\frac{N_s}{\|\mathbf{F}_{\rm RF} \mathbf{V}_{\mathrm{eff}\left(:, 1:N_s \right)}\|_F^2}} \mathbf{V}_{\mathrm{eff}\left(:, 1:N_s \right)},
	\end{align}
    where $\mathbf{V}_{\mathrm{eff}}$ is the right singular matrix of $\mathbf{H}_{\mathrm{eff}}$.
    This satisfies the power constraint of $\| \mathbf{F}_{\mathrm{RF}}\mathbf{F}_{\mathrm{BB}} \|_F^2 = N_s$.
    \subsection{Joint Design of the Phase-shifter Array Pattern and Analog TPC}\label{S4.1}
    In this subsection, we propose a dynamic hybrid TPC method for jointly designing the phase-shifter array pattern and the analog TPC for the dynamic phase-shifter network.
    Note that the proposed method can also be applied in the fixed phase-shifter network.

    According to the above discussion, $\mathbf{F}_{\mathrm{BB}}$ satisfies the power constraint in (\ref{ObjSU}) and $\mathbf{F}_{\mathrm{BB}} \mathbf{F}_{\mathrm{BB}}^{\text{H}} = \gamma \mathbf{I}$, where $\gamma = \frac{N_s}{\| \mathbf{F}_{\mathrm{RF}}\mathbf{V}_{\mathrm{eff}\left(:, 1:N_s \right)} \|_F^2}$.
    Then the optimization problem in (\ref{ObjSU}) is equivalently transformed as
    \begin{subequations}\label{ObjSU_re}
    \begin{equation}
    \max_{\mathbf{F}_{\mathrm{RF}}} \quad \mathrm{log}_2\left(\left| \mathbf{I}_{N_s} + \frac{\rho\gamma}{N_s\sigma^2} \widehat{\mathbf{H}}\mathbf{F}_{\mathrm{RF}} \mathbf{F}_{\mathrm{RF}}^{\text{H}} \widehat{\mathbf{H}}^{\text{H}} \right|\right)
    \end{equation}
    \begin{equation}
    s.t. \quad \left|\mathbf{F}_{\mathrm{RF}}\left(i,j \right)\right| = \frac{1}{\sqrt{N_\mathrm{BS}}},
    \end{equation}
    \end{subequations}
    where $ \widehat{\mathbf{H}} = \mathbf{W}_{\mathrm{opt}}^{\text{H}} \mathbf{H}$.
    It is worth mentioning that this is a non-convex problem due to the non-convex constraints on the entries of $\mathbf{F}_{\mathrm{RF}}$ formulated as $\left|\mathbf{F}_{\mathrm{RF}}\left(i,j \right)\right| = \frac{1}{\sqrt{N_\mathrm{BS}}}$.
    Thus, we propose a entry-by-entry design method for conceiving the joint design of $\mathbf{F}_{\mathrm{RF}}$ and the phase-shifter array pattern.
    We approximate the problem (\ref{ObjSU_re}) in the high-SNR regime as in \cite{8331836}
    \begin{subequations}\label{ObjSU_re2}
    \begin{equation}
    \max_{\mathbf{F}_{\mathrm{RF}}} \quad \mathrm{log}_2\left(\left|\widehat{\mathbf{H}}\mathbf{F}_{\mathrm{RF}} \mathbf{F}_{\mathrm{RF}}^{\text{H}} \widehat{\mathbf{H}}^{\text{H}} \right|\right)
    \end{equation}
    \begin{equation}
    s.t. \quad \left|\mathbf{F}_{\mathrm{RF}}\left(i,j \right)\right| = \frac{1}{\sqrt{N_\mathrm{BS}}}.
    \end{equation}
    \end{subequations}

    We propose to jointly design the phase-shifter array pattern and the quantized phases of the entries in $\mathbf{F}_{\mathrm{RF}}$.
    Specifically, $\mathbf{F}_{\mathrm{RF}}$ is initialized as $\mathbf{F}_{\mathrm{RF}} = \mathbf{F}_{\mathrm{opt}}$ \cite{8295113}, where $\mathbf{F}_{\mathrm{opt}}$ is composed of the $N_s$ largest right singular vectors of the channel matrix $\mathbf{H}$, i.e., $\mathbf{F}_{\mathrm{opt}}=\mathbf{V}_{\left(:, 1:N_s \right)}$.
    Note that $\mathbf{F}_{\mathrm{opt}}$ is only used for initialization and calculation, while the finite-resolution phase shifters are configured according to the quantized phases in $\mathbf{F}_{\rm RF}$.
    Once the index sets $\mathcal{S}_{\mathrm{H}}$ and $\mathcal{S}_{\mathrm{L}}$ are determined, the corresponding phase-shifter array pattern is acquired.
    Thus, we reformulate the problem of designing the phase-shifter array pattern as the design of the index sets.
    For our dynamic twin-resolution phase-shifter network, we propose to derive first the low-resolution phase-shifter array pattern  $\mathcal{S}_{\mathrm{L}}$.
    Then $\mathcal{S}_{\mathrm{H}}$ is acquired as the complementary set of $\mathcal{S}_{\mathrm{L}}$.
    We initialize $\mathcal{S}_{\mathrm{L}}$ to be an empty set, i.e., $\mathcal{S}_{\mathrm{L}} = \varnothing$.

    In order to strike a performance vs. complexity trade-off, we propose to design $\mathcal{S}_{\mathrm{L}}$ and the low-resolution phase shifters in $\mathcal{S}_{\mathrm{L}}$ on a column by column basis.
    Specifically, we divide $\mathcal{S}_{\mathrm{L}}$ into $N_s$ subsets as $\mathcal{S}_{\mathrm{L}} = \{ \mathcal{S}_{\mathrm{L}}^1, \mathcal{S}_{\mathrm{L}}^2, \cdots, \mathcal{S}_{\mathrm{L}}^{N_s} \}$, where $\mathcal{S}_{\mathrm{L}}^i$ denotes the index set of the low-resolution phase shifters in the $i$-th group of phase shifters (i.e. in the $i$-th column of the phase-shifter array pattern).
    We will introduce the design of the first subset $\mathcal{S}_{\mathrm{L}}^1$ and the corresponding first column in the analog TPC matrix.
    Then we can adopt the same procedure for the subsequent design.
    We calculate the minimum quantization error of the phases $\{\varphi_{i,1}\}_{i=1}^{N_{\mathrm{BS}}}$ of the entries, which are not designed  in the first column of $\mathbf{F}_{\mathrm{RF}}$ as follows
    \begin{align}\label{q_l_min}
    q^{{\mathrm{L}},\mathrm{min}}_{i,1} = \arg\min_{q^{\mathrm{L}} \in \mathcal{Q}_{\mathrm{L}}} \left| \varphi_{i,1} - q^{\mathrm{L}} \right|, i\in \{ 1,2,\cdots,N_{\mathrm{BS}}\}\backslash \mathcal{S}_{\mathrm{L}}^1,
	\end{align}
    where $q^{\rm L}$ denotes an arbitrary element in the discrete low-resolution phase shifter set $\mathcal{Q}_{\mathrm{L}}$.
    Then we can acquire the element $q^{{\mathrm{L}},\mathrm{min}}_{i,1}$ in $\mathcal{Q}_{\mathrm{L}}$, which has the minimum quantization error for a certain phase $\varphi_{i,1}$.
    Meanwhile, we can determine the index $\left( i_1,1 \right)$ for the phase $\varphi_{i_1,1}$, which has the minimum quantization error among all entries that are not quantized.
    The index is expressed by
    \begin{align}\label{i11}
    \left( i_1,1 \right) = \arg\min_{i} \left| \varphi_{i,1} - q^{\mathrm{L},\mathrm{min}}_{i,1} \right|.
	\end{align}
    Then we update $\mathcal{S}_{\mathrm{L}}^1$ as $\mathcal{S}_{\mathrm{L}}^1 = \mathcal{S}_{\mathrm{L}}^1 \cup \{ \left( i_1 , 1 \right) \}$.
    Meanwhile, we normalize the amplitude of $\mathbf{F}_{\mathrm{RF}}\left( i_1,1 \right)$ as $\frac{1}{\sqrt{N_\mathrm{BS}}}$ and quantize the phase as $q^{{\mathrm{L}},\mathrm{min}}_{i_1,1}$, which denotes the corresponding quantized phase for $\varphi_{i_1,1}$.
    Thus, we can express the updated entry in $\mathbf{F}_{\mathrm{RF}}$ as
    \begin{align}\label{F_i11_Q}
    \mathbf{F}_{\mathrm{RF}}\left( i_1,1 \right)=\frac{1}{\sqrt{N_{\mathrm{BS}}}}e^{j q^{{\mathrm{L}},\mathrm{min}}_{i_1,1}}.
	\end{align}
    Afterwards, the procedures in (\ref{q_l_min})-(\ref{F_i11_Q}) are repeated until both the joint design of $\mathcal{S}_{\mathrm{L}}^1$ and the updating of the entries in the first column of $\mathbf{F}_{\mathrm{RF}}$ is accomplished.
    We point out that $\{\varphi_{i,1}\}_{i=1,i\notin \mathcal{S}_{\mathrm{L}}^1}^{N_{\mathrm{BS}}}$ in (\ref{q_l_min}) are replaced by $\{\varphi^{\max}_{i,1}\}_{i=1,i\notin \mathcal{S}_{\mathrm{L}}^1}^{N_{\mathrm{BS}}}$ during the iterative process and we further rewrite (\ref{q_l_min}) as
    \begin{align}\label{q_l_minmax}
    q^{{\mathrm{L}},\mathrm{min}}_{i,1} = \arg\min_{q^{\mathrm{L}} \in \mathcal{Q}_{\mathrm{L}}} \left| \varphi_{i,1}^{\max} - q^{\mathrm{L}} \right|, i\in \{ 1,2,\cdots,N_{\mathrm{BS}}\}\backslash \mathcal{S}_{\mathrm{L}}^1,
	\end{align}
    where $\{\varphi^{\max}_{i,1}\}_{i=1,i\notin \mathcal{S}_{\mathrm{L}}^1}^{N_{\mathrm{BS}}}$ are calculated by the associated quantized entries and the other initial entries in $\mathbf{F}_{\mathrm{RF}}$ for maximizing (\ref{ObjSU_re2}a).
    Similarly, $\varphi_{i,1}$ in (\ref{i11}) is changed to $\varphi_{i,1}^{\max}$.
    The process of obtaining $\{\varphi^{\max}_{i,1}\}_{i=1,i\notin \mathcal{S}_{\mathrm{L}}^1}^{N_{\mathrm{BS}}}$ will be discussed later.

    After we complete the design of $\mathcal{S}_{\mathrm{L}}^1$, we can express the index set of low-resolution phase shifters by
    \begin{align}\label{S_L1_set}
    \mathcal{S}_{\mathrm{L}}^1 = \{\left( i_1,1\right), \left( i_2,1\right), \cdots, \left( i_{N_{\mathrm{BS}}/2 }, 1\right) \}.
    \end{align}
    The corresponding updated entries are represented by
    \begin{align}\label{Q_entries_L1}
    \mathbf{F}_{\mathrm{RF}}\left( i,1 \right)= \frac{1}{\sqrt{N_{\mathrm{BS}}}}e^{j q^{{\mathrm{L}},\mathrm{min}}_{i,1}}, \forall \left(i,1\right) \in \mathcal{S}_{\mathrm{L}}^1.
    \end{align}
    Afterwards, we continue to design the remaining subsets of $\mathcal{S}_{\mathrm{L}}$ and the remaining columns of $\mathbf{F}_{\mathrm{RF}}$.
    Finally, we arrive at the index set $\mathcal{S}_{\mathrm{L}}$ of low-resolution phase shifters and the quantized entries formulated as
    \begin{align}\label{Q_entries_L}
    \mathbf{F}_{\mathrm{RF}}\left( i,j \right)= \frac{1}{\sqrt{N_{\mathrm{BS}}}}e^{j q^{{\mathrm{L}},\mathrm{min}}_{i,j}}, \forall \left(i,j\right) \in \mathcal{S}_{\mathrm{L}}.
    \end{align}
    Then a similar method is adopted to design the high-resolution phase shifters given the known $\mathcal{S}_{\mathrm{H}}$, which is the complementary set of $\mathcal{S}_{\mathrm{L}}$.
    For the entry of $\mathbf{F}_{\mathrm{RF}}\left( i,j \right)$ with $\left(i,j\right) \in \mathcal{S}_{\mathrm{H}}$, we quantize the phase of $\mathbf{F}_{\mathrm{RF}}\left( i,j \right)$ as $q^{{\mathrm{H}},\mathrm{min}}_{i,j}$ column by column based on a similar iterative process to that of designing $\{ \mathbf{F}_{\mathrm{RF}}\left( i,j \right) \}_{\left( i,j \right) \in \mathcal{S}_{\mathrm{L}}}$. The difference is that $\mathcal{S}_{\mathrm{L}}$ and $\mathcal{Q}_{\mathrm{L}}$ are replaced by $\mathcal{S}_{\mathrm{H}}$ and $\mathcal{Q_{\mathrm{H}}}$.
    Given that $\mathcal{S}_{\mathrm{H}}$ is acquired, we only have to calculate $ \varphi_{i,j}^{\max}$ for the entries having unquantized phases during the design of the $j$-th column.
    Then we design the specific entry whose phase has the minimum quantization error using a similar procedure to (\ref{q_l_min})-(\ref{F_i11_Q}).
    Finally, we acquire the quantized entries as
    \begin{align}\label{F_ij_Q}
    \mathbf{F}_{\mathrm{RF}}\left( i,j \right)=\left\{
    \begin{aligned}
     &  \frac{1}{\sqrt{N_{\mathrm{BS}}}}e^{j q^{{\mathrm{H}},\mathrm{min}}_{i,j}} , & \left(i,j\right) \in \mathcal{S}_{\mathrm{H}}\\
     &  \frac{1}{\sqrt{N_{\mathrm{BS}}}}e^{j q^{{\mathrm{L}},\mathrm{min}}_{i,j}} , & \left(i,j\right) \in \mathcal{S}_{\mathrm{L}}
    \end{aligned}
    \right.
	\end{align}

    Next, we will propose the calculation method of $\varphi^{\max}_{i,j}$ based on the associated quantized entries and the other initial entries for maximizing (14a), which is an essential intermediate step of the proposed dynamic hybrid TPC method.
    Specifically, we first derive the relationship between (\ref{ObjSU_re2}a) and $\mathbf{F}_{\mathrm{RF}}\left( i,j \right)$.
    Then we partition $\mathbf{F}_{\mathrm{RF}}$ as
    \begin{align}\label{Dec_FRF}
    \mathbf{F}_{\mathrm{RF}} =
    \begin{bmatrix}
    \overline{\mathbf{F}}_{\mathrm{RF}}^{j} & \mathbf{f}_{\mathrm{RF}}^{j}
    \end{bmatrix},
	\end{align}
    where $\overline{\mathbf{F}}_{\mathrm{RF}}^{j} $ denotes the sub-matrix of $\mathbf{F}_{\mathrm{RF}}$ with the arbitrary $j$-th column removed, while $\mathbf{f}_{\mathrm{RF}}^{j}$ represents the arbitrary $j$-th column of $\mathbf{F}_{\mathrm{RF}}$.
    We further rewrite (\ref{ObjSU_re2}a) as
    \begin{align}\label{decomp_1}
    &\mathrm{log}_2\left(\left|\widehat{\mathbf{H}}\mathbf{F}_{\mathrm{RF}} \mathbf{F}_{\mathrm{RF}}^{\text{H}} \widehat{\mathbf{H}}^{\text{H}} \right|\right) \nonumber\\
    \approx & \mathrm{log}_2\left(\left| \mathbf{C}_j \right| \right) + \mathrm{log}_2\left(\left| 1 + \left( \mathbf{f}_{\mathrm{RF}}^{j} \right)^{\text{H}} \mathbf{G}_j \mathbf{f}_{\mathrm{RF}}^{j} \right| \right),
	\end{align}
    where $\mathbf{C}_j = \widehat{\mathbf{H}} \overline{\mathbf{F}}_{\mathrm{RF}}^{j} \left(\overline{\mathbf{F}}_{\mathrm{RF}}^{j}\right)^{\text{H}} \widehat{\mathbf{H}}^{\text{H}}$ and $\mathbf{G}_j = \widehat{\mathbf{H}}^{\text{H}}\left( \xi\mathbf{I}_{Ns} + \right. $ $\left.  \mathbf{C}_j \right)^{-1}\widehat{\mathbf{H}}$ with $\xi$ denoting a small scalar, which guarantees that the matrix inversion exists \cite{6375940}.
    The derivation of (\ref{decomp_1}) is shown in Appendix A.
    Since only the term $\left( \mathbf{f}_{\mathrm{RF}}^{j} \right)^{\text{H}} \mathbf{G}_j \mathbf{f}_{\mathrm{RF}}^{j}$ contains $\mathbf{F}_{\mathrm{RF}}\left( i,j \right)$, we further decompose it as
    \begin{align}\label{ObjSU_re4}
    &\left( \mathbf{f}_{\mathrm{RF}}^{j} \right)^{\text{H}} \mathbf{G}_j \mathbf{f}_{\mathrm{RF}}^{j}\nonumber\\
    = & \left|\mathbf{F}_{\mathrm{RF}}\left( i,j \right)\right| e^{-j\varphi_{i,j}}\sum_{m\neq i}^{N_{\mathrm{BS}}}\left|\mathbf{F}_{\mathrm{RF}}\left( m,j \right)\right| e^{j\varphi_{m,j}} \mathbf{G}_j \left( i,m \right) \nonumber \\
    + & \left|\mathbf{F}_{\mathrm{RF}}\left( i,j \right)\right| e^{j\varphi_{i,j}}\sum_{n\neq i}^{N_{\mathrm{BS}}}\left|\mathbf{F}_{\mathrm{RF}}\left( n,j \right)\right| e^{-j\varphi_{n,j}} \mathbf{G}_j \left( n,i \right)\nonumber  \\
    + & \sum_{n\neq i}^{N_{\mathrm{BS}}} \left|\mathbf{F}_{\mathrm{RF}}\left( n,j \right)\right| e^{-j\varphi_{n,j}} \sum_{m\neq i}^{N_{\mathrm{BS}}} \left|\mathbf{F}_{\mathrm{RF}}\left( m,j \right)\right| e^{j\varphi_{m,j}} \mathbf{G}_j \left( n,m \right) \nonumber \\
    + & \left|\mathbf{F}_{\mathrm{RF}}\left( i,j \right)\right|^2 \mathbf{G}_j \left( i,i \right),
	\end{align}
    where $\varphi_{i,j}$ denotes the phase of $\mathbf{F}_{\mathrm{RF}}\left( i,j \right)$.
    Then we design the phase of $\mathbf{F}_{\mathrm{RF}}\left( i,j \right)$ as $\angle \mathbf{F}_{\mathrm{RF}}\left( i,j \right) = \varphi_{i,j}^{\max}$ that maximizes (\ref{ObjSU_re4}).
    It is plausible that only the first two terms in (\ref{ObjSU_re4}) contain $\varphi_{i,j}$.
    Since the first two terms are each other's conjugate, they have the same magnitudes.
    Then $\varphi_{i,j}^{\max}$ is acquired by the maximization formulated as:
    \begin{align}\label{ObjSU_re5}
    \max_{\varphi_{i,j}}\quad \{e^{-j\varphi_{i,j}}e^{j\varphi_a} + e^{j\varphi_{i,j}}e^{j\varphi_b}\},
	\end{align}
    where $\varphi_a = \angle \sum_{m\neq i}^{N_{\mathrm{BS}}} \left|\mathbf{F}_{\mathrm{RF}}\left( m,j \right)\right| e^{j\varphi_{m,j}} \mathbf{G}_j \left( i,m \right)$ and $\varphi_b = \angle \sum_{n\neq i}^{N_{\mathrm{BS}}}\left|\mathbf{F}_{\mathrm{RF}}\left( n,j \right)\right| e^{-j\varphi_{n,j}} \mathbf{G}_j \left( n,i \right)$.
    Furthermore, due to the fact that $\mathbf{G}_j$ is a Hermitian matrix, we have $\varphi_a = -\varphi_b$.
    Thus, (\ref{ObjSU_re5}) is equal to
    \begin{align}\label{ObjSU_re6}
    \max_{\varphi_{i,j}}\quad \cos\left( \varphi_{i,j} - \varphi_a \right).
	\end{align}
    Finally, we acquire $\varphi_{i,j}^{\max} = \varphi_a$.
    Note that (\ref{ObjSU_re6}) is the solution in the idealized case that the phase shifters have infinite resolution.
    It is simply used for obtaining $\mathbf{F}_{\mathrm{RF}}$ during the proposed dynamic hybrid TPC method.

    We summarize the proposed dynamic hybrid TPC method in Algorithm 1.
    In steps 2-15, we jointly design the phase-shifter array pattern and the phases of the entries in $\mathbf{F}_{\mathrm{RF}}$.
    Finally, we design the digital TPC $\mathbf{F}_{\mathrm{BB}}$ in steps 16-17.

    We point out that the proposed dynamic hybrid TPC method can be slightly modified for employment in the fixed phase-shifter network, when $\mathcal{S}_{\mathrm{H}}$ and $\mathcal{S}_{\mathrm{L}}$ are predetermined.
    Note that in this case, we only have to design the entries in $\mathbf{F}_{\mathrm{RF}}$ according to the fixed pattern.
    We first propose to obtain $\{ \mathbf{F}_{\mathrm{RF}}\left( i,j \right) \}_{\left( i,j \right) \in \mathcal{S}_{\mathrm{L}}}$ column by column.
    Specifically, during the design of the first column in $\mathbf{F}_{\mathrm{RF}}$, we calculate the optimal quantized phases $\{q^{{\mathrm{L}},\mathrm{min}}_{i,1}\}_{i \in {\mathcal{S}_{\mathrm{L}}^1}}$ among the fixed set $\mathcal{S}_{\mathrm{L}}^1$ according to (\ref{q_l_min}).
    Then we update the entry $\mathbf{F}_{\mathrm{RF}}\left( i_1,1 \right)$ according to (\ref{F_i11_Q}).
    We calculate $\{\varphi^{\max}_{\tilde{i},1}\}_{\tilde{i}=1,\tilde{i}\notin \mathcal{S}_{\mathrm{L}}^1}^{N_{\mathrm{BS}}}$ according to (\ref{ObjSU_re6}) for the subsequent design.
    Then the design of the first column can be accomplished by repeating the aforementioned procedure.
    Afterwards, we continue to design $\{ \mathbf{F}_{\mathrm{RF}}\left( i,j \right) \}_{\left( i,j \right) \in \mathcal{S}_{\mathrm{L}} \backslash \mathcal{S}_{\mathrm{L}}^1 }$ for the remaining columns.
    After we complete the design of phase shifters having low resolution, we continue to design $\{ \mathbf{F}_{\mathrm{RF}}\left( i,j \right) \}_{\left( i,j \right) \in \mathcal{S}_{\mathrm{H}}}$ column by column.

    We now proceed with our complexity analysis as follows.
    The computational complexity of our proposed method is composed of two parts, namely the analog TPC design and the digital TPC design.
    The digital TPC is acquired by the SVD of the effective channel $\mathbf{H}_{\mathrm{eff}}$ at a computational complexity order of $\mathcal{O}\left(N_{\rm {RF}}N_s^2\right)$.
    During the analog TPC design, as part of the initialization, we have to compute the SVD of $\mathbf{H}$, which has the computational complexity order of $\mathcal{O}\left(N_{\mathrm{BS}}^3\right)$.
    Then, the calculation of $\mathbf{C}_j$ and $\mathbf{G}_j$ is executed at an order of $\mathcal{O}\left(N_{\mathrm{BS}}\left(N_{\rm {RF}}-1\right)N_s\right)$ and $\mathcal{O}\left(N_{\mathrm{BS}}^2N_s\right)$, respectively, which are repeated $N_{\rm RF}$ times.
    Furthermore, the calculation of phases has the computational complexity order of $\mathcal{O}\left(N_{\mathrm{BS}} - 1\right)$, which is repeated $\frac{N_{\mathrm{BS}}^2N_{\mathrm{RF}}+ N_{\mathrm{BS}}N_{\mathrm{RF}} - 2 N_{\mathrm{BS}}}{2}$ times.
    Therefore, the overall complexity order of the proposed algorithm is $\mathcal{O}\left( N_{\mathrm{BS}}N_{\rm {RF}}^2N_s + N_{\mathrm{BS}}^2N_{\rm RF}N_s + \frac{N_{\rm BS}^3}{2}N_{\rm RF} + N_{\rm BS}^3 + N_{\rm {RF}}N_s^2 \right)$.

\begin{algorithm}[h]
\caption{Proposed Dynamic Twin-Resolution Hybrid TPC for MmWave Point-to-point MIMO Systems}
    \begin{algorithmic}[1]
        \REQUIRE ~~\\
        $\mathbf{F}_{\mathrm{opt}}$, $\mathbf{W}_{\mathrm{opt}}$, $\mathbf{H}$, $\mathcal{S}_{\mathrm{H}}$, $\mathcal{S}_{\mathrm{L}}$, $N_{\mathrm{BS}}$ and $N_s$ ;
        \ENSURE
        \STATE Initialize $\mathbf{F}_{\mathrm{RF}}$ as $\mathbf{F}_{\mathrm{opt}}$;
        \FOR{$j = 1 : N_s$}
        \STATE Calculate $\mathbf{C}_{j}$, $\mathbf{G}_{j}$;
        \FOR{$i = 1 : \frac{N_{\mathrm{BS}}}{2}$}
        \STATE Jointly update the index set $\mathcal{S}_{\mathrm{L}}^j$ according to (\ref{q_l_min})-(\ref{i11}) and design ${\{ \mathbf{F}_{\mathrm{RF}}}\left( i,j \right) \}_{\left( i,j \right) \in \mathcal{S}_{\mathrm{L}}}$ as $\frac{1}{\sqrt{N_{\mathrm{BS}}}}e^{j q^{{\mathrm{L}},\mathrm{min}}_{i,j}}$;
        \STATE Calculate $\{\varphi^{\max}_{\tilde{i},j}\}_{\tilde{i}=1,\tilde{i}\notin \mathcal{S}_{\mathrm{L}}^j}^{N_{\mathrm{BS}}}$ according to (\ref{ObjSU_re6});
        \ENDFOR
        \ENDFOR
        \FOR{$j = 1 : N_s$}
        \STATE Calculate $\mathbf{C}_{j}$, $\mathbf{G}_{j}$;
        \FOR{$ \left(i,j\right) \in \mathcal{S}_{\mathrm{H}}$}
        \STATE Design $\{ {\mathbf{F}_{\mathrm{RF}}} \left( i,j \right) \}_{\left( i,j \right) \in \mathcal{S}_{\mathrm{H}}}$ as $ \frac{1}{\sqrt{N_{\mathrm{BS}}}}e^{j q^{{\mathrm{H}},\mathrm{min}}_{i,j}}$;
        \STATE Calculate $\varphi^{\max}_{\tilde{i},j}$ for the entries whose phases have not been quantized in the $j$-th column.
        \ENDFOR
        \ENDFOR
        \STATE Calculate the effective channel $\mathbf{H}_{\mathrm{eff}} = \mathbf{W}_{\mathrm{opt}}^{\text{H}} \mathbf{H} \mathbf{F}_{\mathrm{RF}}$;
        \STATE {Calculate $\mathbf{F}_{\mathrm{BB}} = \sqrt{\frac{N_s}{\|\mathbf{F}_{\rm RF} \mathbf{V}_{\mathrm{eff}\left(:, 1:N_s \right)}\|_F^2}} \mathbf{V}_{\mathrm{eff}\left(:, 1:N_s \right)}$.}
        \label{code:recentEnd}
    \end{algorithmic}
\end{algorithm}
    \subsection{Performance Analysis}\label{S4.2}
        In this subsection, we will characterize the performance difference between the fully digital solution and the proposed method.
    We sequentially replace the entries of the fully digital solution with those of the hybrid TPC matrix derived.
    We calculate the bandwidth efficiency variation in each replacement step, where the sum of the bandwidth efficiency changes represents the overall bandwidth efficiency gap between the fully digital solution and the proposed method.
    {By exploiting the equation $\log_2\left( \mathbf{I} + \mathbf{X}\mathbf{Y} \right) = \log_2\left( \mathbf{I} + \mathbf{Y}\mathbf{X} \right)$,
    we equivalently express the bandwidth efficiency as
    \begin{align}\label{SU_SE_HP}
    \mathbf{\textit{R}} &= \mathrm{log}_2 \left( \left| \mathbf{I}_{N_s} + \frac{\rho}{N_s\sigma^2} \mathbf{W}_{\mathrm{opt}}^{\text{H}} \mathbf{H} \mathbf{F}_{\mathrm{RF}} \mathbf{F}_{\mathrm{BB}} \mathbf{F}^{\text{H}}_{\mathrm{BB}} \mathbf{F}^{\text{H}}_{\mathrm{RF}} \mathbf{H}^{\text{H}} \mathbf{W}_{\mathrm{opt}} \right|\right)\nonumber \\
    & = \mathrm{log}_2\left(\left| \mathbf{I}_{N_s} + \frac{\rho}{N_s\sigma^2} \mathbf{F}_{\mathrm{BB}}^{\text{H}}\mathbf{F}^{\text{H}}_{\mathrm{RF}} \mathbf{M} \mathbf{F}_{\mathrm{RF}}\mathbf{F}_{\mathrm{BB}}\right|\right)\nonumber \\
    & = \mathrm{log}_2\left(\left| \mathbf{I}_{N_s} + \frac{\rho}{N_s\sigma^2} \mathbf{F}^{\text{H}} \mathbf{M} \mathbf{F}\right|\right)
	\end{align}
    where $\mathbf{M}=\mathbf{H}^{\text{H}} \mathbf{W}_{\mathrm{opt}} \mathbf{W}_{\mathrm{opt}}^{\text{H}} \mathbf{H}$, $\mathbf{F} = \mathbf{F}_{\mathrm{RF}}\mathbf{F}_{\mathrm{BB}}$ denotes our hybrid TPC matrix designed.}
    Then we sequentially replace the entries of $\mathbf{F}_{\mathrm{opt}}$ by that of $\mathbf{F}$ for deriving the bandwidth efficiency gap between the designed $\mathbf{F}$ and full-digital SVD solution.
    We denote the TPC matrix before and after the $k$-th replacement by $\mathbf{F}^{\left( k - 1 \right)}$ and $\mathbf{F}^{\left( k \right)}$, respectively.
    {In the $k$-th replacement, the $\left( i,j \right)$-th element in $\mathbf{F}^{\left( k - 1 \right)}$ is replaced by the corresponding element with the index $\left( i,j\right)$ in the hybrid TPC matrix $\mathbf{F}$ derived to generate $\mathbf{F}^{\left( k \right)}$, where $k = \left( j-1 \right)N_{\rm{RF}} + i$.
    We define $\mathbf{f}^{j}$ as the $j$-th column of $\mathbf{F}$ and $\overline{\mathbf{F}}^{j}$ as the submatrix of $\mathbf{F}$ composed of the residual columns.
    We rewrite (\ref{SU_SE_HP}) as (\ref{SU_SE_HP2}).}
    \begin{figure*}[t]
    {\begin{align}\label{SU_SE_HP2}
    \mathbf{\textit{R}} = & \mathrm{log}_2\left(\left| \mathbf{I}_{N_s} + \frac{\rho}{N_s\sigma^2} \mathbf{F}^{\text{H}} \mathbf{M} \mathbf{F}\right|\right) \nonumber \\
    = &  \mathrm{log}_2\left(\left| \mathbf{I}_{N_s} + \frac{\rho}{N_s\sigma^2} \begin{bmatrix} \overline{\mathbf{F}}^{j} & \mathbf{f}^j \end{bmatrix}^{\text{H}} \mathbf{M} \begin{bmatrix} \overline{\mathbf{F}}^{j} & \mathbf{f}^j \end{bmatrix}\right|\right) \nonumber \\
    = & \mathrm{log}_2\left(\left|
    \begin{bmatrix} \mathbf{I}_{N_s-1} + \frac{\rho}{N_s\sigma^2} \left( \overline{\mathbf{F}}^{j} \right)^{\text{H}}\mathbf{M} \overline{\mathbf{F}}^{j} & \frac{\rho}{N_s\sigma^2} \left( \overline{\mathbf{F}}^{j} \right)^{\text{H}}\mathbf{M} \mathbf{f}^{j} \\
    \frac{\rho}{N_s\sigma^2} \left( \mathbf{f}^{j} \right)^{\text{H}}\mathbf{M} \overline{\mathbf{F}}^{j} & 1 + \frac{\rho}{N_s\sigma^2} \left( \mathbf{f}^{j} \right)^{\text{H}}\mathbf{M} \mathbf{f}^{j}
    \end{bmatrix}\right|\right)
	\end{align}}
    \hrulefill
    \end{figure*}
    {By exploiting the formula
    \begin{align}\label{formula}
    \left| \begin{bmatrix} \mathbf{A}_1 & \mathbf{A}_2\\ \mathbf{A}_3 & \mathbf{A}_4 \end{bmatrix} \right| =\left| \mathbf{A}_1 \right|\left| \mathbf{A}_4 - \mathbf{A}_3 \mathbf{A}_1^{-1}\mathbf{A}_2 \right| ,
	\end{align}
    we can arrive at
    \begin{align}\label{SU_SE_HP3}
    \mathbf{\textit{R}} = & \mathrm{log}_2\left(\left| \mathbf{D}_j  \right| \left( 1 + \left(\mathbf{f}^{j} \right)^{\text{H}}\mathbf{Y}_j \mathbf{f}^{j} \right) \right) \nonumber\\
    = & \mathrm{log}_2\left(\left| \mathbf{D}_j  \right|\right)  + \mathrm{log}_2\left(  1 + \left(\mathbf{f}^{j} \right)^{\text{H}}\mathbf{Y}_j \mathbf{f}^{j} \right),
	\end{align}
    where $\mathbf{D}_j = \mathbf{I}_{N_s-1} + \frac{\rho}{N_s\sigma^2} \left( \overline{\mathbf{F}}^{j}\right)^{\text{H}}\mathbf{M}\overline{\mathbf{F}}^{j}$, $\mathbf{Y}_j = \frac{\rho}{N_s\sigma^2}\mathbf{M} - \frac{\rho^2}{N_s^2\sigma^4}\mathbf{M}\overline{\mathbf{F}}^{j} \mathbf{D}_{j}^{-1} \left( \overline{\mathbf{F}}^{j} \right)^{\text{H}} \mathbf{M}$.
    Afterwards, by adopting the equation in Theorem 2 of \cite{6364537}, we decompose (\ref{SU_SE_HP3}) into its element-wise representation as}
    \begin{align}\label{SU_HPSE_Column}
    \mathbf{\textit{R}} = \mathrm{log}_2\left(\left| \mathbf{D}_j \right|\right) + \mathrm{log}_2\left( 1 + e_{i,j} + f_{i,j} + g_{i,j} \right),
	\end{align}
    where
%    and $\overline{\mathbf{F}}_{\mathrm{RF}}^{j} $ denotes the sub-matrix of $\mathbf{F}_{\mathrm{RF}}$ with the $j$-th column removed.
    \begin{align}\label{eHP}
    e_{i,j} = \sum_{m\neq i}^{N_{\mathrm{BS}}}\sum_{n\neq i}^{N_{\mathrm{BS}}}\mathbf{F}^{\ast}\left( m,j \right) \mathbf{Y}_j\left( m,n \right)\mathbf{F}\left( n,j \right),
	\end{align}
    \begin{align}\label{fHP}
    f_{i,j} & = {2 \mathfrak{Re}\left\{ \left( \sum_{m\neq i}^{N_{\mathrm{BS}}} \mathbf{F}^{\ast}\left( m,j \right) \mathbf{Y}_j\left( m,i \right) \right) \mathbf{F}\left( i,j \right)  \right\}} \nonumber\\
    & = 2 \left| \mathbf{F}\left( i,j \right) \right|\left| b_{i,j} \right| \mathrm{cos}\left( \psi_{i,j} + a_{ij} \right),
	\end{align}
    with $a_{ij} = \angle \mathbf{F}\left( i,j \right)$, $b_{ij} = \sum_{m\neq i}^{N_{\mathrm{BS}}} \mathbf{F}^{\ast}\left( m,j \right) \mathbf{Y}_j\left( m,i \right)$ and $\psi_{i,j} = \angle b_{i,j}$,
    \begin{align}\label{gHP}
    g_{i,j} = \frac{\mathbf{Y}_j\left( i,i \right)}{\left|\mathbf{F}\left( i,j \right)\right|}.
	\end{align}
    Therefore, the performance loss in the $k$-th replacement is expressed as
    \begin{align}\label{R_delta}
    R_{\Delta}^{\left( k \right)}= \mathrm{log}_2\left( \frac{ 1 + e_{i,j}^{\left( k \right)} + f_{i,j}^{\left( k \right)} + g_{i,j}^{\left( k \right)} }{ 1 + e_{i,j}^{\left( k - 1 \right)} + f_{i,j}^{\left( k - 1 \right)} + g_{i,j}^{\left( k - 1 \right)} } \right),
	\end{align}
    where $e_{i,j}^{\left( k \right)}$, $f_{i,j}^{\left( k \right)}$ and $g_{i,j}^{\left( k \right)} $ denote the expression of $e_{i,j}$, $f_{i,j}$ and $g_{i,j}$ in the case when $\mathbf{F}^{\left( k \right)}$ is substituted into (\ref{SU_HPSE_Column}).
    Note that we have $e_{i,j}^{\left( k - 1 \right)} = e_{i,j}^{\left( k \right)}$ due to the fact that $e_{i,j}$ is not related to the $\left( i,j \right)$-th entry of $\mathbf{F}$.
    Then we may transform the performance loss $R_{\Delta}^{\left( k \right)}$ of Eq. (\ref{R_delta}) into
    \begin{align}\label{R_delta1}
    R_{\Delta}^{\left( k \right)} &= \mathrm{log}_2\left( \frac{ 1 + e_{i,j}^{\left( k - 1 \right)} + f_{i,j}^{\left( k \right)} + g_{i,j}^{\left( k \right)} }{ 1 + e_{i,j}^{\left( k - 1 \right)} + f_{i,j}^{\left( k - 1 \right)} + g_{i,j}^{\left( k - 1 \right)} } \right)  \nonumber \\
    & = \mathrm{log}_2\left( 1 + \frac{ \Delta f^{\left( k \right)} + \Delta g^{\left( k \right)} }{ 1 + e_{i,j}^{\left( k - 1 \right)} + f_{i,j}^{\left( k - 1 \right)} + g_{i,j}^{\left( k - 1\right)} } \right) ,
	\end{align}
    where we have
    \begin{align}\label{delta_fk}
    \Delta f^{\left( k \right)} & = f_{i,j}^{\left( k \right)} - f_{i,j}^{\left( k - 1 \right)} \nonumber \\
    & = 2 \left| b^{\left( k-1 \right)}_{i,j} \right| \left( \left| \mathbf{F}^{\left( k \right)}\left( i,j \right) \right|\mathrm{cos}\left( \psi^{\left( k-1 \right)}_{i,j} + a_{ij}^{k}  \right) \right.\nonumber\\
    & \left. -  \left| \mathbf{F}^{\left( k-1 \right)}\left( i,j \right) \right|\mathrm{cos}\left( \psi^{\left( k-1 \right)}_{i,j} + a_{ij}^{k-1} \right)  \right),
    \end{align}
    \begin{align}\label{delta_gk}
    \Delta g^{\left( k \right)} & = g_{i,j}^{\left( k \right)} - g_{i,j}^{\left( k - 1 \right)}\nonumber\\
    &= \mathbf{Y}^{\left( k - 1 \right)}_j\left( i,i \right)\left( \frac{1}{\left|\mathbf{F}^{\left( k \right)}\left( i,j \right)\right|} - \frac{1}{\left|\mathbf{F}^{\left( k - 1 \right)}\left( i,j \right)\right|} \right),
    \end{align}
    {where $b_{i,j}^{\left( k - 1 \right)}$, $\psi_{i,j}^{\left( k - 1 \right)}$ and $\mathbf{Y}^{\left( k - 1 \right)}_j\left( i,i \right)$ do not change because they are not related to the $\left( i,j \right)$-th entry of $\mathbf{F}$.}
    Furthermore, the bandwidth efficiency gap between the SVD solution and the proposed dynamic hybrid TPC method is expressed as
    \begin{align}\label{R_deltaSVD}
    R_{\Delta} = \sum_{k=1}^{N_{\mathrm{BS}}N_s}R_{\Delta}^{\left( k \right)}.
	\end{align}
    \section{Proposed Dynamic Hybrid Precoding Method for MmWave MU-MIMO Systems}\label{S5}
    Similar to Section \ref{S4}, we design the combiner for each user as the optimal SVD solution.
    The SVD of the channel $\mathbf{H}_u$ is given by $\mathbf{H}_u = \mathbf{U}_u\mathbf{\Sigma}_u\mathbf{V}_u^{\text{H}}$.
    Then the combiner between the BS and the $u$-th user is designed as $\mathbf{W}_{\mathrm{opt},u}=\mathbf{U}_{u\left(:, 1:M_s \right)}$.
    The bandwidth efficiency of the $u$-th user is rewritten as
    \begin{align}\label{R_u_re}
    \mathbf{\emph{R}}_u & =  \mathrm{log}_2 \left( \left| \mathbf{I}_{N_s} + {\frac{\rho}{N_s}} \mathbf{R}_{\mathrm{opt},i}^{-1} \mathbf{W}_{\mathrm{opt},u}^{\text{H}} \mathbf{H}_u \mathbf{F}_{\mathrm{RF}} \mathbf{F}_{\mathrm{BB},u} \right.\right. \nonumber \\ & \quad \quad \quad \quad \quad \left.\left.\mathbf{F}_{\mathrm{BB},u}^{\text{H}} \mathbf{F}^{\text{H}}_{\mathrm{RF}} \mathbf{H}_u^{\text{H}} \mathbf{W}^u_{\mathrm{opt}} \right|\right),
	\end{align}
    where $\mathbf{R}_{\mathrm{opt},i} = \sum_{i=1,i \neq u}^U\mathbf{W}_{\mathrm{opt},u}^{\text{H}} \mathbf{H}_u \mathbf{F}_{\mathrm{RF}} \mathbf{F}_{\mathrm{BB},i} \mathbf{F}_{\mathrm{BB},i}^{\text{H}}$ $  \mathbf{F}^{\text{H}}_{\mathrm{RF}}\mathbf{H}_u^{\text{H}}\mathbf{W}_{\mathrm{opt},u} + \sigma^2\mathbf{I}_{M_s}$.

    We first present the design of the baseband TPC for eliminating the inter-user interference and the interference among the data streams of each user, when the analog TPC $\mathbf{F}_{\mathrm{RF}}$ is given.
    Afterwards, we extend the proposed dynamic hybrid TPC method to the mmWave MU-MIMO systems.

    The effective channel between the BS and the $u$-th user is expressed as $\mathbf{H}_{\mathrm{eff},u} = \mathbf{W}_{\mathrm{opt},u}^{\text{H}} \mathbf{H}_u \mathbf{F}_{\mathrm{RF}} \in \mathbb{C}^{M_s \times N_s} $.
    Then we adopt the classic block diagonalization method for eliminating the inter-user interference and the interference among data streams for each user, which is given by \cite{7335586}
    \begin{align}\label{FBB_MU}
    {\widetilde{\mathbf{F}}_{\mathrm{BB},u}}  = \overline{\mathbf{V}}_{u \left[:,\left( U-1 \right)M_s+1:UM_s\right]}\widetilde{\mathbf{V}}_{u \left[:, 1:M_s\right]},
	\end{align}
    where $\overline{\mathbf{V}}_{u}$ is composed of the right singular vectors of $\overline{\mathbf{H}}_{\mathrm{eff},u} = \left[ \left( \mathbf{H}_{\mathrm{eff},1} \right)^{\text{H}}, \cdots, \left( \mathbf{H}_{\mathrm{eff},u-1} \right)^{\text{H}}, \left( \mathbf{H}_{\mathrm{eff},u+1} \right)^{\text{H}} , \cdots , \left( \mathbf{H}_{\mathrm{eff},U} \right)^{\text{H}} \right]^{\text{H}}$ $ \in \mathbb{C}^{\left( U-1 \right)M_s \times N_s}$, which can be expressed as $\overline{\mathbf{H}}_{\mathrm{eff},u} = \overline{\mathbf{U}}_{u}\overline{\mathbf{\Sigma}}_{u}\overline{\mathbf{V}}_{u}^{\text{H}}$.
    The matrix $\widetilde{\mathbf{V}}_{u}$ is composed of the right singular vectors of  $\mathbf{H}_{\mathrm{eff},u}\overline{\mathbf{V}}_{u \left[:,\left( U-1 \right)M_s+1:UM_s\right]}\in \mathbb{C}^{M_s \times M_s}$, whose SVD is expressed as $\mathbf{H}_{\mathrm{eff},u}\overline{\mathbf{V}}_{u \left[:,\left( U-1 \right)M_s+1:UM_s\right]} = \widetilde{\mathbf{U}}_u\widetilde{\mathbf{\Sigma}}_u\widetilde{\mathbf{V}}_u^{\text{H}}$.

    We observe that the right $M_s$ columns of $\overline{\mathbf{V}}_{u}$, i.e., $\overline{\mathbf{V}}_{u \left[:,\left( U-1 \right)M_s+1:UM_s\right]}$, exhibit the property that $\mathbf{H}_{\mathrm{eff},i}\overline{\mathbf{V}}_{u \left[:,\left( U-1 \right)M_s+1:UM_s\right]} = 0$ for $i \neq u$, meaning that the inter-user interference is eliminated.
    Then we have $\mathbf{R}_{\mathrm{opt},i} = \sigma^2\mathbf{I}_{M_s} $.
    We normalize the digital TPC at the $u$-th user as $\mathbf{F}_{\mathrm{BB},u} = {\sqrt{\frac{N_s}{\| \mathbf{F}_{\mathrm{RF}}\widetilde{\mathbf{F}}_{\mathrm{BB},u} \|_F^2}}\widetilde{\mathbf{F}}_{\mathrm{BB},u}}$ to satisfy the power constraint.

    Furthermore, the analog TPC design problem is transformed as
    \begin{subequations}\label{ObjMU}
    \begin{equation}
    \max_{\mathbf{F}_{\mathrm{RF}}} \quad \sum_{u=1}^U \mathrm{log}_2\left(\left| \mathbf{I}_{N_s} + {\frac{\rho}{N_s\sigma^2}} \mathbf{H}_{\mathrm{eff},u} \mathbf{F}_{\mathrm{BB},u} \mathbf{F}_{\mathrm{BB},u}^{\text{H}} \mathbf{H}_{\mathrm{eff},u}^{\text{H}} \right|\right)
    \end{equation}
    \begin{equation}
    s.t.  \quad \left|\mathbf{F}_{\mathrm{RF}}\left(i,j \right)\right| = \frac{1}{\sqrt{N_\mathrm{BS}}}.
    \end{equation}
    \end{subequations}
    Note that the proposed dynamic hybrid TPC method cannot be directly adopted for solving (\ref{ObjMU}) due to the fact that $\mathbf{F}_{\mathrm{BB},u} \mathbf{F}_{\mathrm{BB},u}^{\text{H}} \neq \mathbf{I}_{M_s}$.
    To solve this problem, we embark on further derivations.
    We substitute the digital TPC into the OF and exploit the property of matrix $\widetilde{\mathbf{V}}_{u}$ that $\widetilde{\mathbf{V}}_{u \left[:, 1:M_s\right]}\widetilde{\mathbf{V}}_{u \left[:, 1:M_s\right]}^{\text{H}} =\mathbf{I}_{M_s}$.
    The OF in (\ref{ObjMU}) can be rewritten as
    \begin{align}\label{SE_FReStep1}
    \sum_{u=1}^U \mathrm{log}_2\left(\left| \mathbf{I}_{M_s} + {\frac{\rho \gamma_u}{N_s\sigma^2}} \mathbf{H}_{\mathrm{eff},u} \left( \overline{\mathbf{V}}_{u} \overline{\mathbf{V}}_{u}^{\text{H}} - \overline{\mathbf{V}}_{u}^{\mathrm{R}}\right) \mathbf{H}_{\mathrm{eff},u} ^{\text{H}} \right|\right) ,
	\end{align}
    where $\gamma_u = \frac{N_s}{\| \mathbf{F}_{\mathrm{RF}}{\widetilde{\mathbf{F}}_{\mathrm{BB},u}} \|_F^2}$ is the coefficient that satisfies the power constraint, and $\overline{\mathbf{V}}_{u}^{\mathrm{R}} = \overline{\mathbf{V}}_{u \left[:,1:\left( U-1 \right)M_s\right]}$ $\overline{\mathbf{V}}_{u \left[:,1:\left( U-1 \right)M_s\right]}^{\text{H}}$.
    Afterwards, we exploit the property of $\overline{\mathbf{V}}_{u}$ that $\overline{\mathbf{V}}_{u}\overline{\mathbf{V}}_{u }^{\text{H}} =\mathbf{I}_{N_s}$ and further rewrite (\ref{SE_FReStep1}) as
    \begin{align}\label{SE_FReStep2}
    & \quad \sum_{u=1}^U \mathrm{log}_2\left(\left| \mathbf{I}_{M_s} + {\frac{\rho \gamma_u}{N_s\sigma^2}}  \mathbf{H}_{\mathrm{eff},u}  \mathbf{H}_{\mathrm{eff},u}^{\text{H}} - \right.\right. \nonumber\\
    & \quad \quad \quad \quad \quad \quad \quad \quad \quad\left.\left. \frac{\rho \gamma_u}{M_s\sigma^2} \mathbf{H}_{\mathrm{eff},u} \overline{\mathbf{V}}_{u}^{\mathrm{R}}  \mathbf{H}_{\mathrm{eff},u}^{\text{H}} \right|\right)\nonumber \\
    & = \sum_{u=1}^U \left( \mathrm{log}_2\left(\left| \mathbf{T}_u \right|\right) + \right.\nonumber\\
    & \quad \quad \quad  \quad \quad  \left. \mathrm{log}_2\left(\left| \mathbf{I}_{M_s} -  \mathbf{T}_u^{-1} {\frac{\rho \gamma_u}{N_s\sigma^2}} \mathbf{H}_{\mathrm{eff},u} \overline{\mathbf{V}}_{u}^{\mathrm{R}} \mathbf{H}_{\mathrm{eff},u}^{\text{H}} \right|\right) \right)\nonumber \\
    & = R_{\mathrm{ideal}} + R_{\mathrm{loss}} ,
	\end{align}
    where $\mathbf{T}_u = \mathbf{I}_{M_s} + \frac{\rho \gamma_u}{\sigma^2} \mathbf{H}_{\mathrm{eff},u} \left( \mathbf{H}_{\mathrm{eff},u} \right)^{\text{H}}$,
    $R_{\mathrm{ideal}} = \sum_{u=1}^U \mathrm{log}_2\left(\left| \mathbf{T}_u \right|\right)$ represents the ideal case that no inter-user interference exists.
    Furthermore, $R_{\mathrm{loss}} = \sum_{u=1}^U \mathrm{log}_2\left(\left| \mathbf{I}_{M_s} - \mathbf{T}_u^{-1} {\frac{\rho \gamma_u}{N_s\sigma^2}} \mathbf{H}_{\mathrm{eff},u} \overline{\mathbf{V}}_{u}^{\mathrm{R}} \left( \mathbf{H}_{\mathrm{eff},u} \right)^{\text{H}} \right|\right)$ represents the performance loss due to the inter-user interference.

    {In this paper, we extend the proposed dynamic hybrid TPC method to MU-MIMO systems by maximizing $R_{\mathrm{ideal}}$, which can be expressed as
\begin{subequations}\label{subObjMURe1}
    \begin{equation}
    \max_{\mathbf{F}_{\mathrm{RF}}} \quad \sum_{u=1}^U \left( \mathrm{log}_2\left(\left| \mathbf{I}_{M_s} + {\frac{\rho \gamma_u}{N_s\sigma^2}} \mathbf{H}_{\mathrm{eff},u} \left( \mathbf{H}_{\mathrm{eff},u} \right)^{\text{H}} \right|\right) \right)
    \end{equation}
    \begin{equation}
    s.t.  \quad \left|\mathbf{F}_{\mathrm{RF}}\left(i,j \right)\right| = \frac{1}{\sqrt{N_\mathrm{BS}}}.
    \end{equation}
    \end{subequations}}
    Specifically, we partition $\mathbf{F}_{\mathrm{RF}}$ into $U$ submatrices, yielding $\mathbf{F}_{\mathrm{RF}} = \left[\mathbf{F}_{\mathrm{RF},1}, \mathbf{F}_{\mathrm{RF},2},\cdots, \mathbf{F}_{\mathrm{RF},U} \right]$.
    We propose to divide the design into $U$ steps.
    In the $u$-th step, we utilize the proposed hybrid TPC method to design the submatrix $\mathbf{F}_{\mathrm{RF},u}$.
    Thus, the subproblem in each step is given by
    \begin{subequations}\label{subObjMURe2}
    \begin{equation}
    \max_{\mathbf{F}_{\mathrm{RF},u}} \quad \mathrm{log}_2\left(\left| \mathbf{I}_{M_s} + {\frac{\rho \gamma_u}{N_s\sigma^2} } \mathbf{H}_{\mathrm{eff},u} \left( \mathbf{H}_{\mathrm{eff},u} \right)^{\text{H}} \right|\right)
    \end{equation}
    \begin{equation}
    s.t.  \quad \left|\mathbf{F}_{\mathrm{RF},u}\left(i,j \right)\right| = \frac{1}{\sqrt{N_\mathrm{BS}}}.
    \end{equation}
    \end{subequations}
   { We then exploit the relationship of $\log_2\left( \mathbf{I} + \mathbf{X}\mathbf{Y} \right) = \log_2\left( \mathbf{I} + \mathbf{Y}\mathbf{X} \right)$ and approximate the problem in the high-SNR regime as}
\begin{subequations}\label{subObjMU}
    \begin{equation}
    \max_{\mathbf{F}_{\mathrm{RF},u}} \quad \mathrm{log}_2\left(\left|\widehat{\mathbf{H}}_u \mathbf{F}_{\mathrm{RF},u}  \mathbf{F}_{\mathrm{RF},u}^{\text{H}}  \widehat{\mathbf{H}}_u^{\text{H}} \right|\right)
    \end{equation}
    \begin{equation}
    s.t.  \quad \left|\mathbf{F}_{\mathrm{RF},u}\left(i,j \right)\right| = \frac{1}{\sqrt{N_\mathrm{BS}}},
    \end{equation}
    \end{subequations}
    where $\widehat{\mathbf{H}}_u= \mathbf{W}_{\mathrm{opt},u}^{\text{H}} \mathbf{H}_u$.

    It is observed that the problem of hybrid TPC design in mmWave MU-MIMO systems is transformed into $U$ subproblems, each of which has the same form representing mmWave point-to-point MIMO systems.
    Therefore, the proposed dynamic hybrid TPC can be adopted to accomplish the design.
    The initial RF TPC is set as $\mathbf{F}_{\mathrm{RF}} = \left[ \mathbf{F}_{\mathrm{opt},1}, \mathbf{F}_{\mathrm{opt},2}, \cdots, \mathbf{F}_{\mathrm{opt},U}\right]$, where $\mathbf{F}_{\mathrm{opt},u} = \mathbf{V}_{u\left(:, 1:M_s \right)}$.
    In the $u$-th step, we jointly design the phase-shifter array pattern and the quantized phases of the entries in $\mathbf{F}_{\mathrm{RF},u}$ using the same procedure as in (\ref{q_l_min}) - (\ref{F_ij_Q}).
%    \subsection{Performance Analysis for MU-MIMO systems}\label{S5.3}
    Similar to the performance analysis of mmWave point-to-point MIMO systems, we present the bandwidth efficiency gap between the fully digital solution $\mathbf{F}_{\mathrm{opt},u}$ and our proposed dynamic hybrid TPC method.
    We define $\mathbf{F}_u = \mathbf{F}_{\mathrm{RF}}\mathbf{F}_{\mathrm{BB},u}$.
    The RF TPC matrix before and after the $k$-th replacement in the $u$-th step are denoted by $\mathbf{F}_{\mathrm{RF}}^{\left(u, k - 1 \right)}$ and $\mathbf{F}_{\mathrm{RF}}^{\left(u, k \right)}$, respectively.
    We substitute them into (\ref{SU_HPSE_Column})-(\ref{delta_gk}) and acquire the bandwidth efficiency difference $R_{\Delta}^{\left( u,k \right)}$ during a single replacement.
    Furthermore, the bandwidth efficiency gap between the SVD-based fully digital solution and the hybrid TPC method is expressed as:
    \begin{align}\label{R_deltaSVD_MU}
    R_{\Delta} = \sum_{u=1}^{U}\sum_{k=1}^{N_{\mathrm{BS}}N_s}R_{\Delta}^{\left( u,k \right)}.
	\end{align}
%    \section{Complexity Analysis}\label{S6}

    {For the MU-MIMO case, calculating the baseband TPC requires the SVD of the effective channels $\left\{\overline{\mathbf{H}}_{\rm{eff},u}\right\}_{u=1}^U$ and $\left\{ \mathbf{H}_{\mathrm{eff},u}\overline{\mathbf{V}}_{u \left[:,\left( U-1 \right)M_s+1:UM_s\right]} \right\}_{u=1}^U$.
    The complexity is on the order of $\mathcal{O}\left[U\left(U-1\right)M_s^3\right]$ and $UM_s^3$, respectively.
    The design of the analog TPC in MU-MIMO systems is accomplished by executing the proposed method in point-to-point MIMO systems $U$ times.
    Thus, the overall complexity of the proposed algorithm for MU-MIMO systems is on the order of $\mathcal{O}\left( UN_{\mathrm{BS}}M_s^3 +  UN_{\mathrm{BS}}^2M_s^2 + \frac{UN_{\rm BS}^3N_{\rm RF}}{2} + UN_{\rm BS}^3 + U^2M_s^3\right. $ $\left. + UM_s^3 \right)$.}
    \section{Numerical Results}\label{S6}
    In this section, we provide simulation results for the proposed hybrid TPC design relying on our twin-resolution phase shifters in both mmWave point-to-point MIMO systems and MU-MIMO systems.
    \subsection{Simulation Results of MmWave Point-to-point MIMO Systems}\label{S6.1}
    Firstly, we consider mmWave point-to-point MIMO systems.
    The transmitter has $N_{\mathrm{BS}} = 8\times8$ TAs and the receiver has $N_{\mathrm{MS}}=4\times4$ RAs.
    The number of RF chains at the BS is set to $N_{\mathrm{RF}} = 4$.
    The number of data streams $N_s$ is set to 4.
    The number of propagation paths is 8.
    The azimuth AoD $\phi^{\mathrm{r}}_{\ell}$ is uniformly distributed in the interval $\left[ -\pi,\pi \right)$ and elevation AoD $\theta^{\mathrm{r}}_{\ell}$ is uniformly distributed in the interval $\left[ -\frac{\pi}{2},\frac{\pi}{2} \right)$ \cite{TWC_AAhmed_LimitedHybridPrecoding}.

\begin{figure}[t]
		\center{\includegraphics[width=0.48\textwidth]{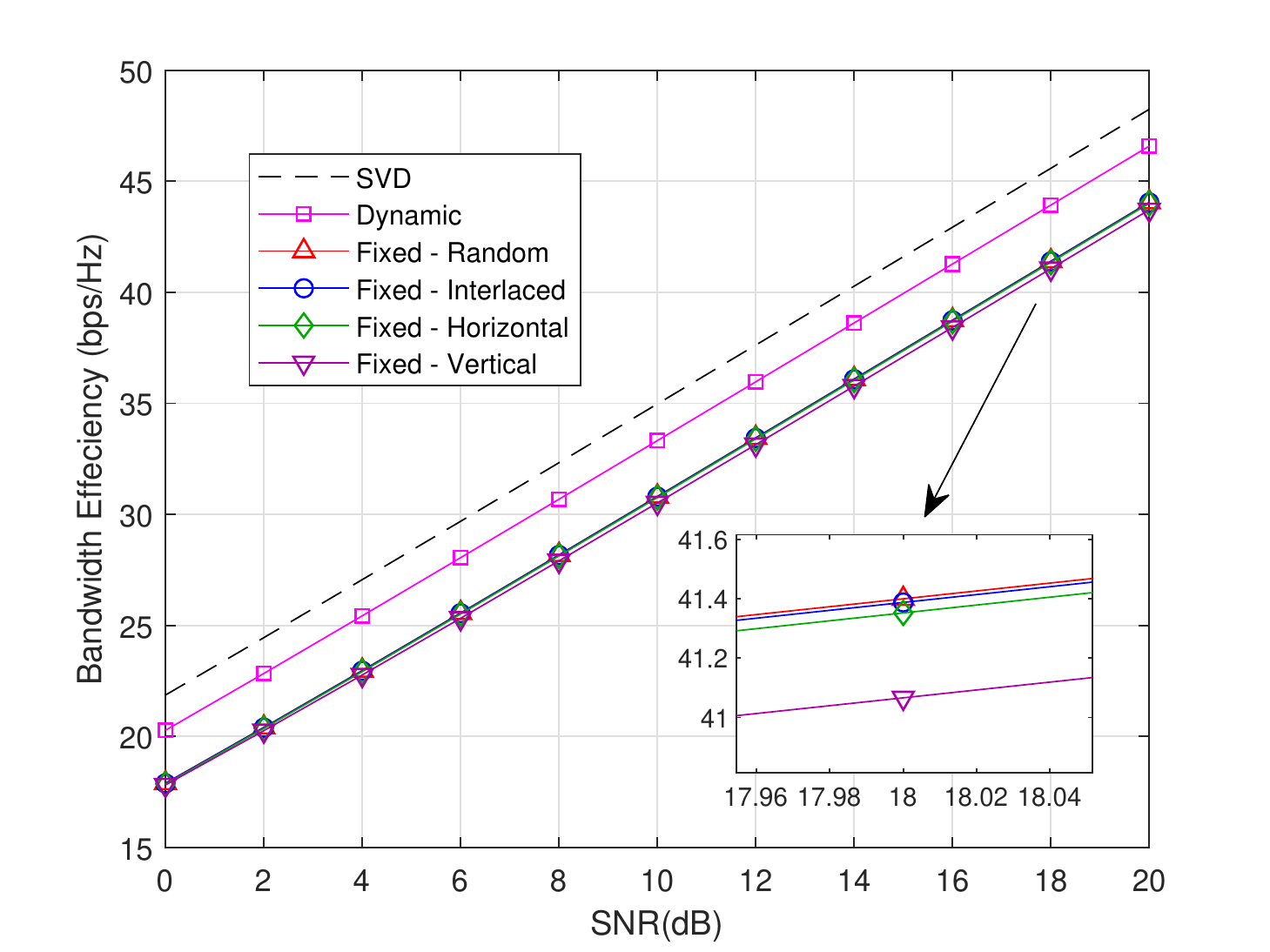}}
		\caption{Bandwidth efficiency of the different phase-shifter networks. $N_{\mathrm{BS}}$ = 64, $N_{\mathrm{MS}}$ = 16, $N_{\mathrm{RF}}$ = 4, $N_s$ = 4.}
		\label{structure_comparison}
	\end{figure}

    We compare the bandwidth efficiency of different phase-shifter networks in Fig. \ref{structure_comparison}.
    Our simulation results illustrate that the random, interlaced and horizontal fixed networks have similar performance and they outperform the vertical fixed network.
    As expected, the dynamic phase-shifter network outperforms the fixed networks, hence, we adopt the dynamic twin-resolution phase-shifter network in the following simulations.

\begin{figure}[t]
\centering
\includegraphics[width=0.45\textwidth]{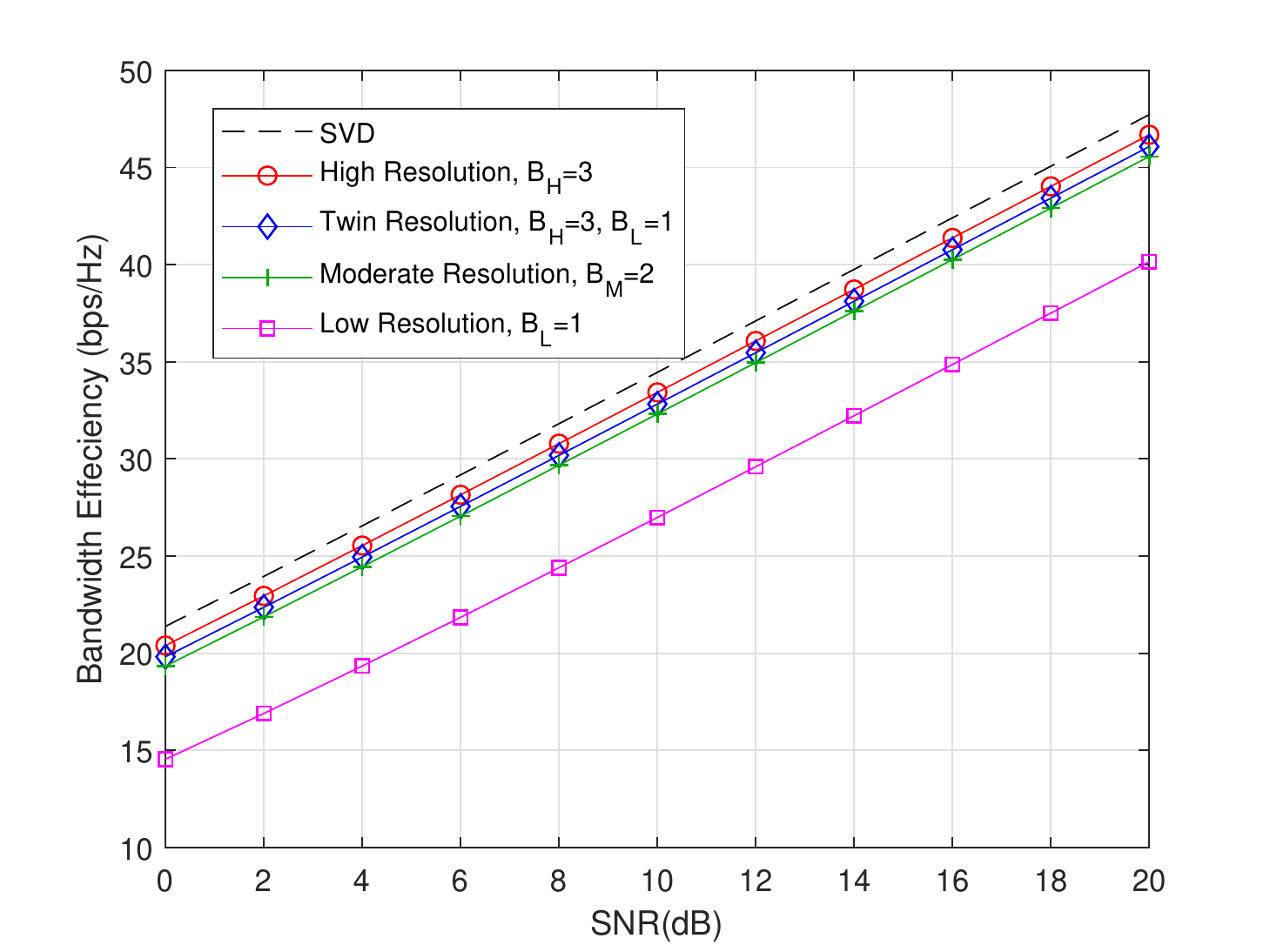}
\caption{Bandwidth efficiency vs. SNR in mmWave point-to-point MIMO systems. $N_{\mathrm{BS}}$ = 64, $N_{\mathrm{MS}}$ = 16, $N_{\mathrm{RF}}$ = 4, $N_s$ = 4.}
\label{SE_vs_SNR_SU}
\end{figure}

    Fig. \ref{SE_vs_SNR_SU} shows the bandwidth efficiency versus SNR.
    We simulate the bandwidth efficiency in 5 cases.
    The high resolution $B_{\mathrm{H}}$ is set to 3 bits while the low resolution $B_{\mathrm{L}}$ is set to 1 bit in the simulations.
    For comparison, we plot the bandwidth efficiency of the entirely high-resolution phase-shifter network and of the entirely low-resolution phase-shifter network.
    Note that the index set $\mathcal{S}_{\mathrm{L}}$ is an empty set and $\mathcal{S}_{\mathrm{H}}$ contains all entries in the analog TPC matrix, when an entirely high-resolution phase-shifter network is adopted.
    Furthermore, the fully digital SVD solution is also shown in Fig. \ref{SE_vs_SNR_SU}.
    It is observed that the twin-resolution phase-shifter network achieves a bandwidth efficiency close to that of the entirely high-resolution phase-shifter network.
    As expected, the twin-resolution phase-shifter network also outperforms the entirely low-resolution phase-shifter network.
    We also plot the bandwidth efficiency of the moderate-resolution phase-shifter network, where the moderate resolution $B_{\mathrm{M}}$ is set to 2 bits.
    As illustrated in Fig. 7, the proposed twin-resolution phase-shifter network has a higher bandwidth efficiency than the moderate-resolution one.

\begin{figure}[t]
\centering
\includegraphics[width=0.45\textwidth]{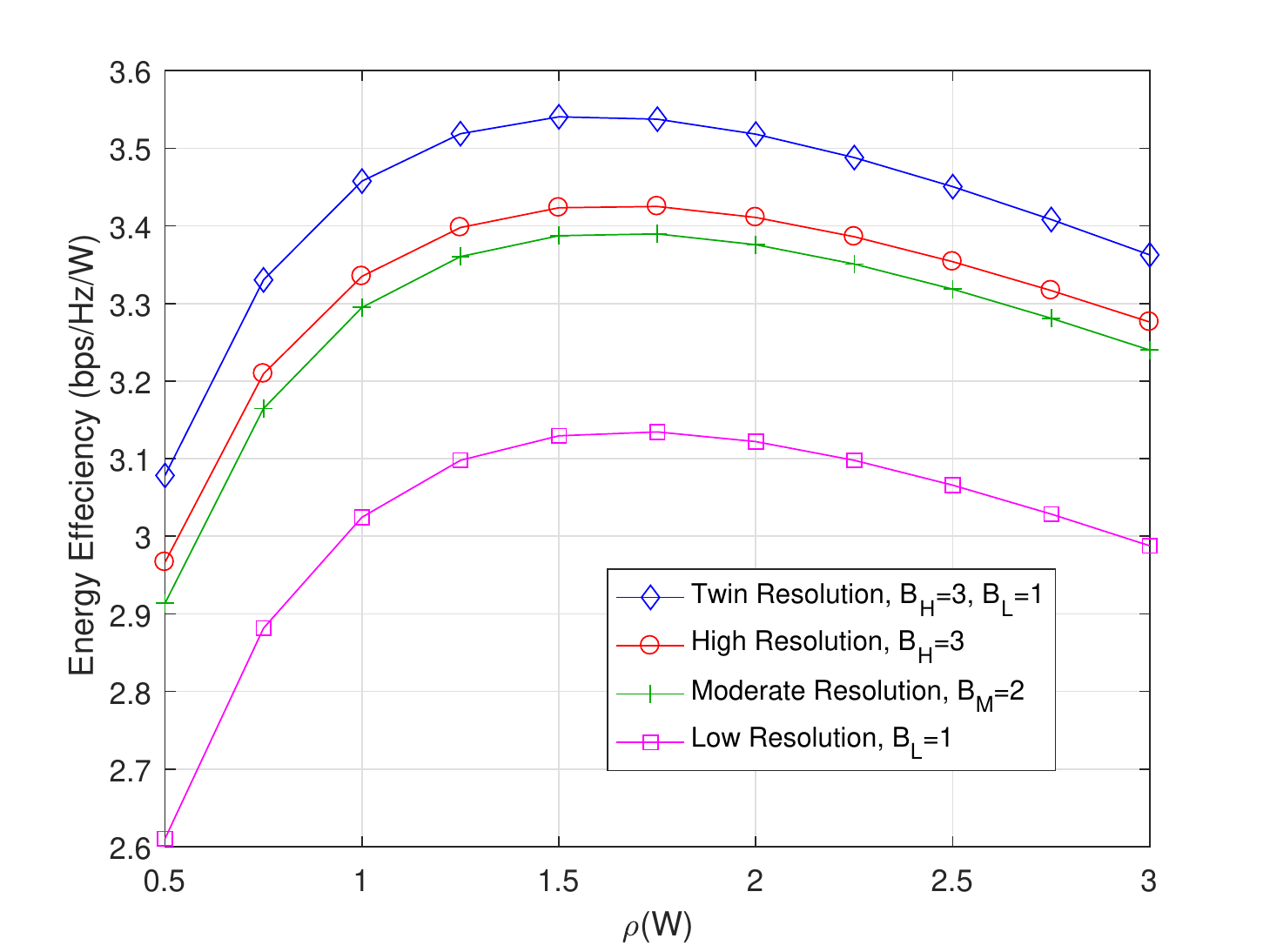}
\caption{Energy efficiency vs. $\rho$ in mmWave point-to-point MIMO systems. $N_{\mathrm{BS}}$ = 64, $N_{\mathrm{MS}}$ = 16, $N_{\mathrm{RF}}$ = 4, $N_s$ = 4, $\sigma^2$ = 1.}
\label{EE_vs_Ps_SU}
\end{figure}

    Our energy efficiency comparison is illustrated in Fig. \ref{EE_vs_Ps_SU}.
    The power consumption of the high-resolution phase shifter is $P_h = 15$mW \cite{7370753}, that of the moderate-resolution phase shifter is $P_m = 14$mW \cite{6291764} and that of the 1-bit low-resolution phase shifter is $P_l = 10$mW \cite{7997065}.
    The power consumption of switches is $P_{\mathrm{SW}}=1$mW \cite{7542170}.
    For the twin-resolution phase-shifter network, the energy efficiency of our twin-resolution phase shifter network is defined as
    \begin{align}\label{EE}
    & \mathrm{EE}_{\mathrm{TR}}\nonumber \\
    = & \frac{R}{\rho + P_{\mathrm{BB}} + N_sP_{\mathrm{RF}} + \frac{N_{\mathrm{BS}}N_s}{2}P_h + \frac{N_{\mathrm{BS}}N_s}{2}P_l + N_{\mathrm{BS}} N_s P_{\mathrm{SW}} },
	\end{align}
    where $P_{\mathrm{BB}} = 250$mW and $P_{\mathrm{RF}} = 300$mW \cite{7997065} are the power consumption of the baseband processor and of a single RF chain.
    Similarly, the energy efficiency of entirely high-resolution, moderate-resolution or 1-bit low-resolution phase-shifter network is given by
    \begin{align}\label{EE_o}
    \mathrm{EE}_{o} = \frac{R}{\rho + P_{\mathrm{BB}} + N_sP_{\mathrm{RF}} + N_{\mathrm{BS}}N_sP_o},
	\end{align}
    where $o \in \{h,m,l\}$ denotes the high-, moderate- or low-resolution phase shifters.
    The noise power is set to $\sigma^2 = 1$.
    The transmit power $\rho$ ranges from 0.5W to 3W.
    It is observed that although the bandwidth efficiency is a little bit lower than that of the entirely high-resolution phase-shifter network, the twin-resolution phase-shifter network is more energy-efficient.
    In Fig. 8, we also observe that the energy efficiency of the twin-resolution phase-shifter network is higher than that of the moderate-resolution network.
\begin{figure}[t]
\centering
\includegraphics[width=0.45\textwidth]{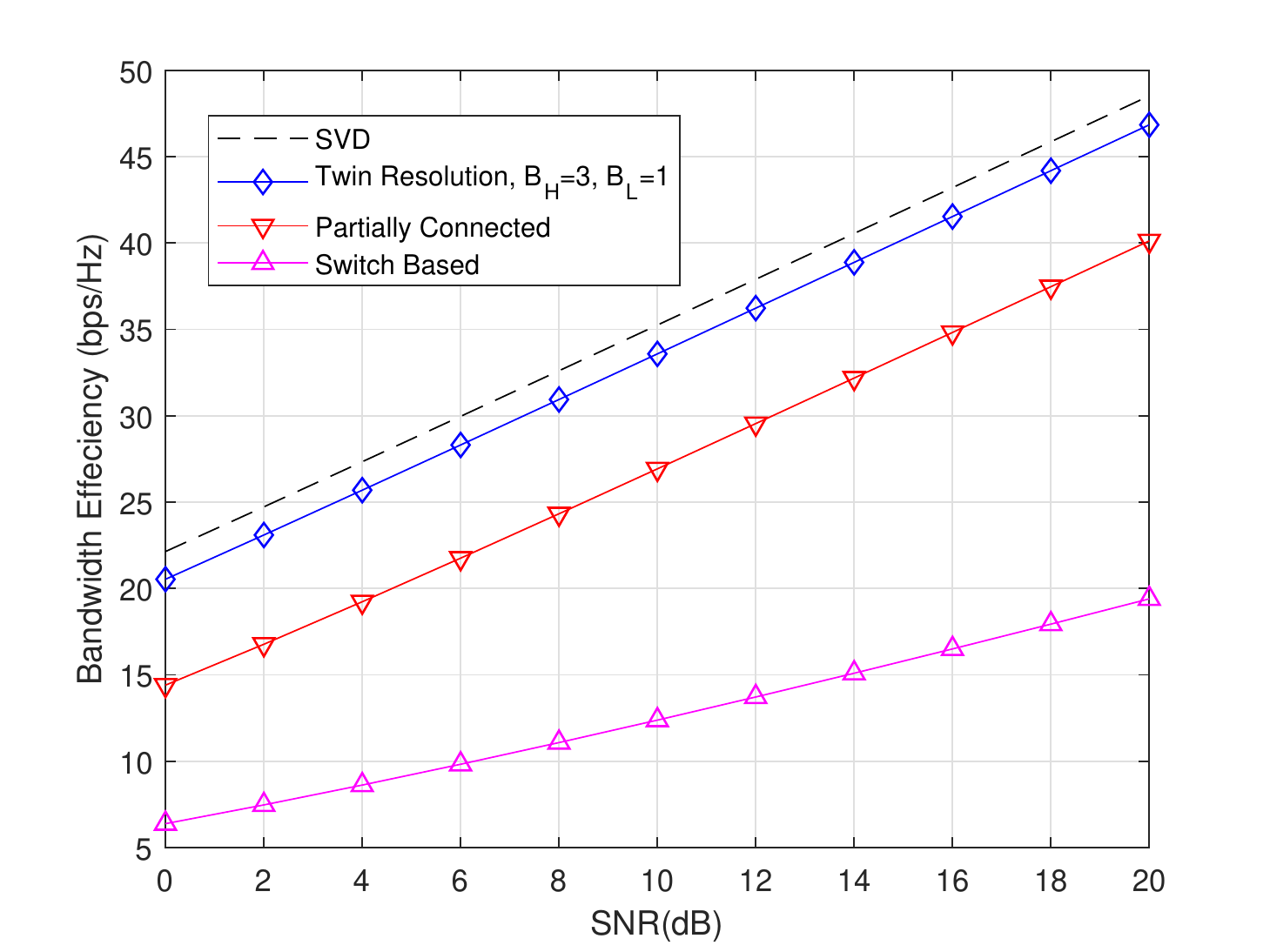}
\caption{{Bandwidth efficiency of different systems. $N_{\mathrm{BS}}$ = 64, $N_{\mathrm{MS}}$ = 16, $N_{\mathrm{RF}}$ = 4, $N_s$ = 4.}}
\label{SE_comp_sys}
\end{figure}
\begin{figure}[t]
\centering
\includegraphics[width=0.45\textwidth]{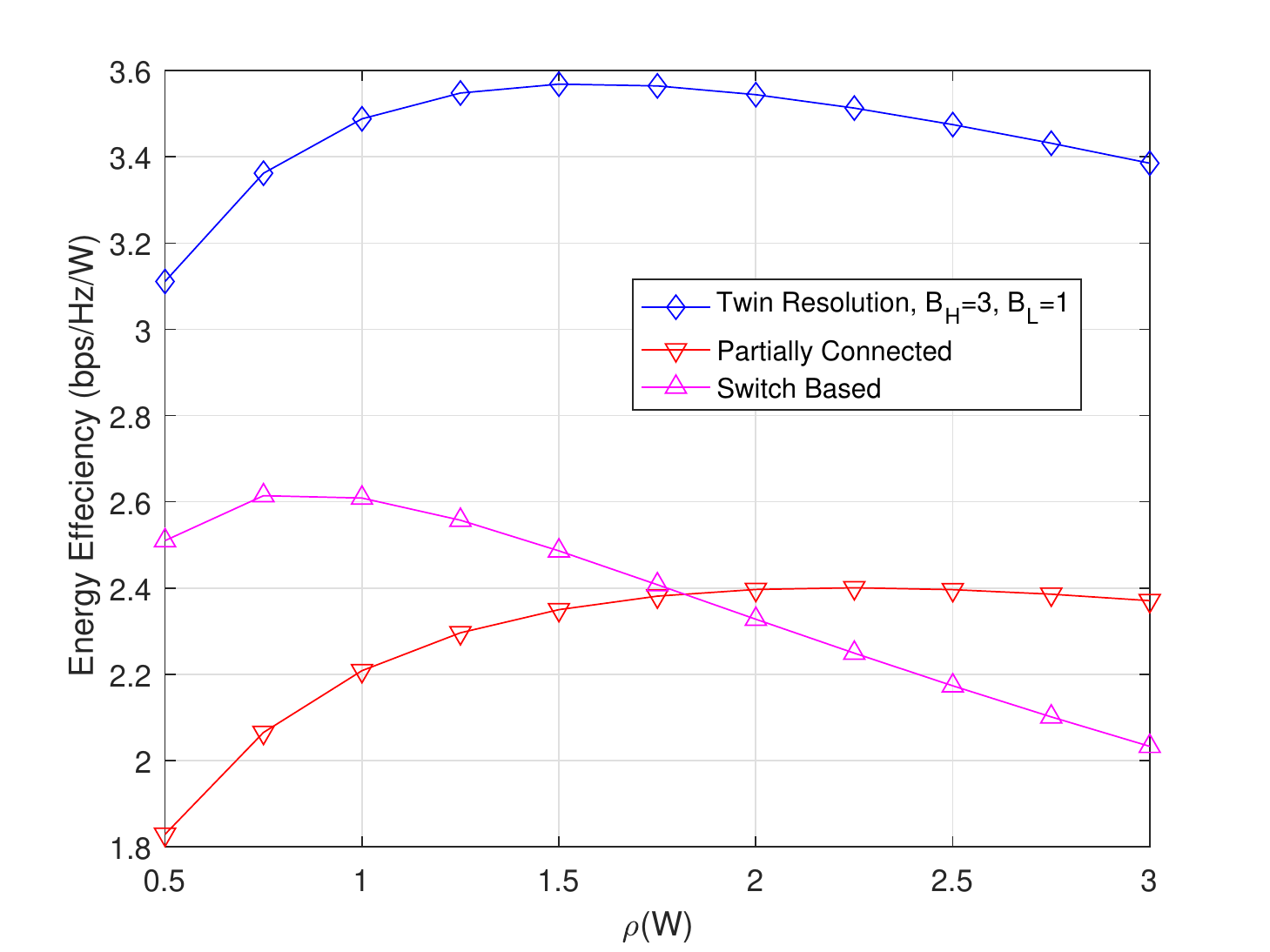}
\caption{{Energy efficiency of different systems. $N_{\mathrm{BS}}$ = 64, $N_{\mathrm{MS}}$ = 16, $N_{\mathrm{RF}}$ = 4, $N_s$ = 4, $\sigma^2$ = 1.}}
\label{EE_comp_sys}
\end{figure}

    We compare the bandwidth efficiency and energy efficiency of our proposed twin-resolution phase-shifter network based system and the other two low-complexity hybrid TPC systems, namely the switch-based system \cite{7370753} and the partially-connected system \cite{7445130}.
In switch-based systems, the low-complexity switch-based network is adopted for analog TPC.
The OMP-based method is adopted for designing the hybrid TPC of the switch-based systems.
In the partially-connected phase-shifter network, each RF chain is connected to a single sub-array having $\frac{N_{\rm BS}}{N_{\rm RF}}$ antennas.
The successive interference cancellation (SIC)-based method is adopted for designing the hybrid TPC in the partially-connected systems.
For energy efficiency comparison, we set the resolution of the phase shifters in partially-connected network to 3 bits.
All other parameters are the same as those in Fig. \ref{SE_vs_SNR_SU} and Fig. \ref{EE_vs_Ps_SU}.
As shown in Fig. \ref{SE_comp_sys} and Fig. \ref{EE_comp_sys}, our proposed twin-resolution phase-shifter network based system outperforms both the switch-based system and the partially-connected system both in terms of its bandwidth efficiency and energy efficiency.

    \subsection{Simulation Results of MmWave MU-MIMO Systems}\label{S6.2}
\begin{figure}[t]
\centering
\includegraphics[width=0.45\textwidth]{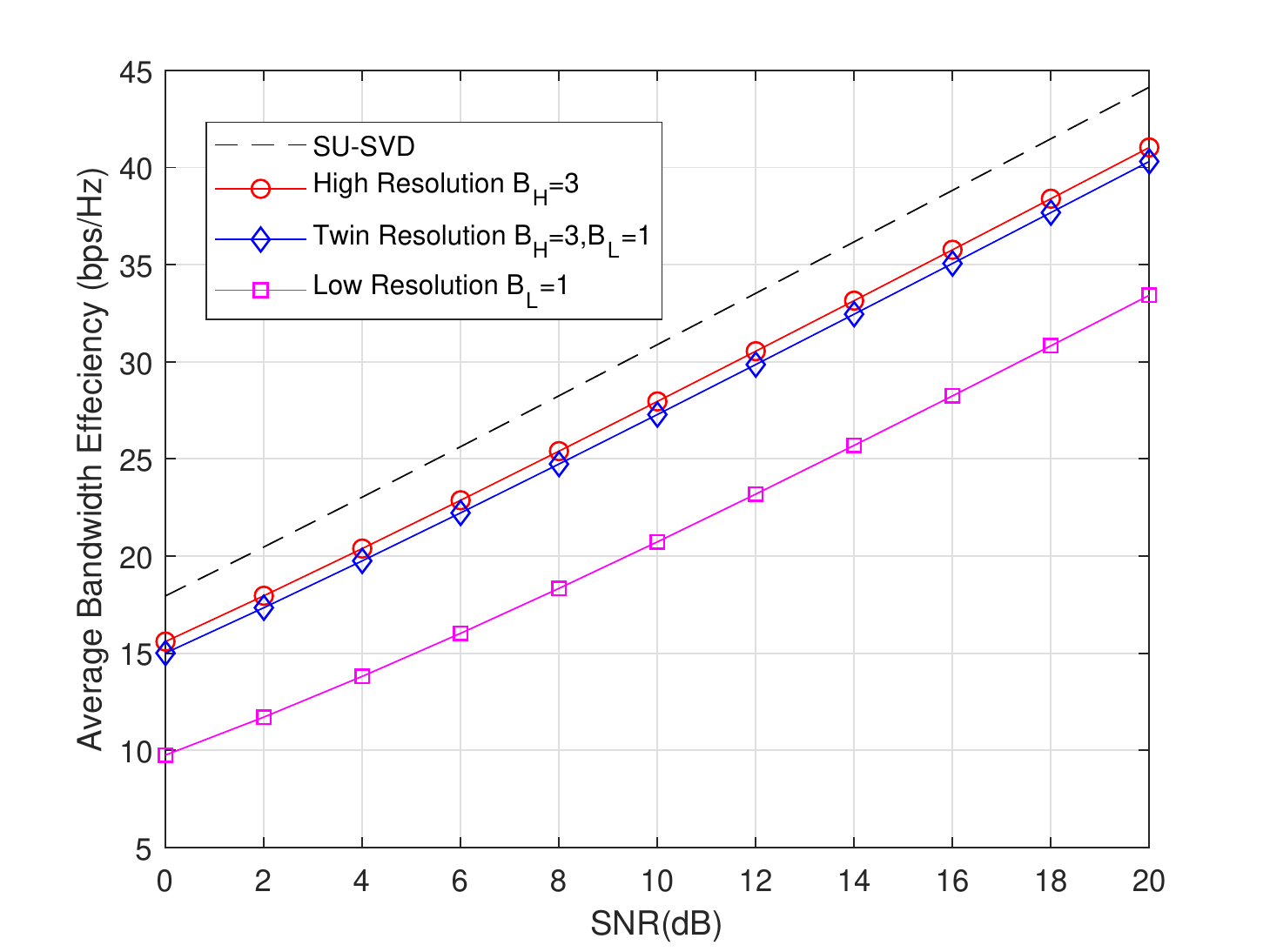}
\caption{Average bandwidth efficiency vs. SNR in MU-MIMO systems. $N_{\mathrm{BS}}$ = 64, $N_{\mathrm{MS}}$ = 16, $U$ = 2, $M_s$ = 4, $N_{\mathrm{RF}}$ = 8, $N_s$ = 8.}
\label{SE_vs_SNR_MU}
\end{figure}
\begin{figure}[t]
\centering
\includegraphics[width=0.45\textwidth]{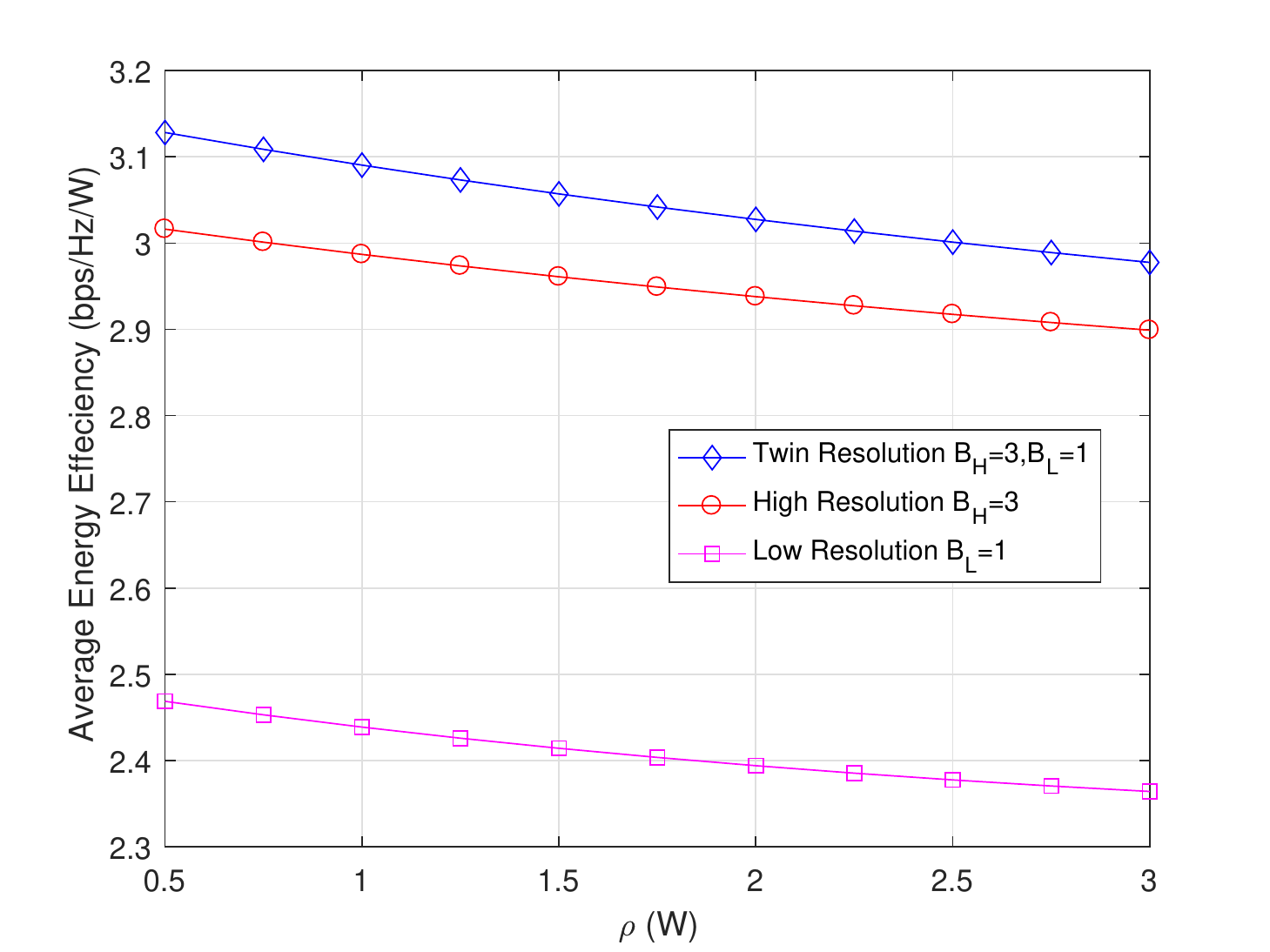}
\caption{Energy efficiency vs. $\rho$ in MU-MIMO systems. $N_{\mathrm{BS}}$ = 64, $N_{\mathrm{MS}}$ = 16, $U$ = 2, $M_s$ = 4, $N_{\mathrm{RF}}$ = 8, $N_s$ = 8, $\sigma^2$ = 1.}
\label{EE_vs_Ps_MU}
\end{figure}
\begin{figure}[t]
\centering
\includegraphics[width=0.45\textwidth]{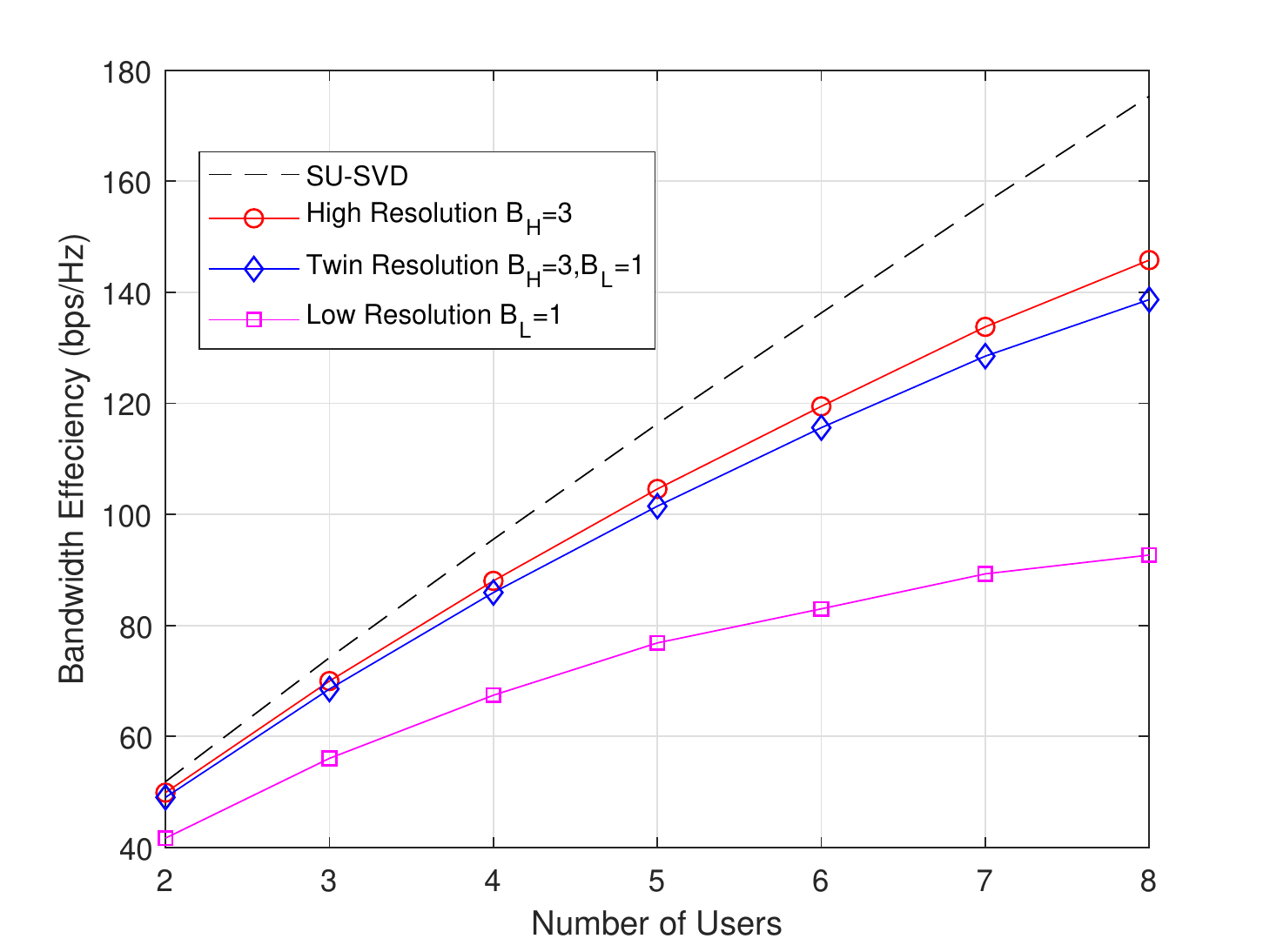}
\caption{Bandwidth efficiency vs. the number of users in MU-MIMO systems. $N_{\mathrm{BS}}$ = 64, $N_{\mathrm{MS}}$ = 16, $M_s$ = 2, $\text{SNR}$ = 20dB.}
\label{SE_vs_U_MU}
\end{figure}
\begin{figure}[t]
\centering
\includegraphics[width=0.45\textwidth]{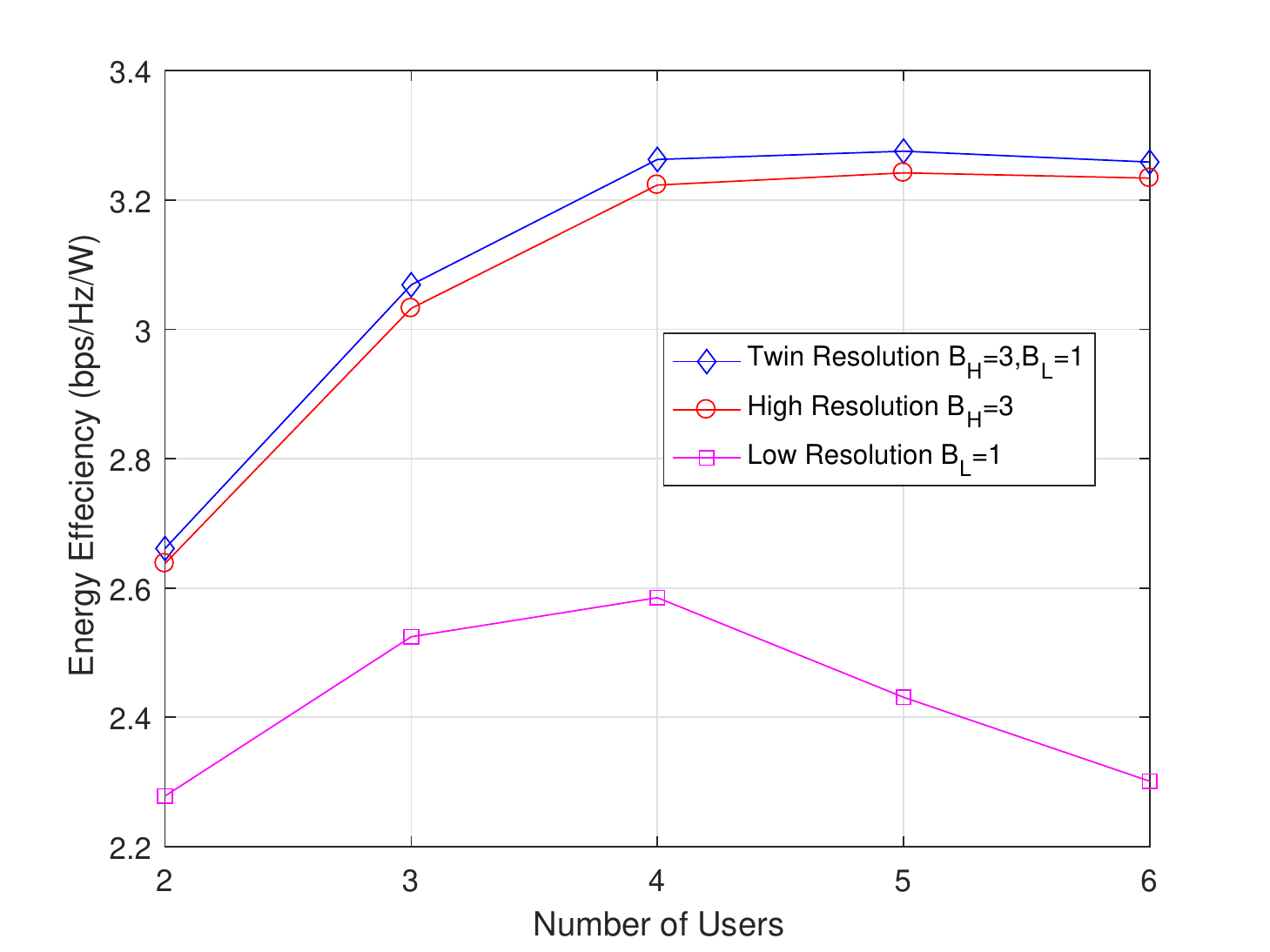}
\caption{Energy efficiency vs. the number of users in MU-MIMO systems. $N_{\mathrm{BS}}$ = 64, $N_{\mathrm{MS}}$ = 16, $M_s$ = 2, $\rho$ = 1W, $\sigma^2$ = 1.}
\label{EE_vs_U_MU}
\end{figure}
    In MU-MIMO systems, the energy efficiency of our twin-resolution phase shifter network is defined as
    \begin{align}\label{EEMU}
    &\mathrm{EE}_{\mathrm{TR-MU}} = \nonumber \\
    & \frac{R}{\rho + P_{\mathrm{BB}} + UM_sP_{\mathrm{RF}} + \frac{UN_{\mathrm{BS}}M_s}{2}\left(P_h + P_l\right) + UN_{\mathrm{BS}} M_s P_{\mathrm{SW}}},
	\end{align}
    and that of the entirely high-resolution or 1-bit low-resolution phase-shifter network is given by
    \begin{align}\label{EE_o}
    EE_{o\mathrm{-MU}} = \frac{R}{\rho + P_{\mathrm{BB}} + UM_sP_{\mathrm{RF}} + UN_{\mathrm{BS}}M_sP_o}.
	\end{align}
    There are 2 users in the system.
    The number of data streams for each user is $M_S = 4$.
    At the BS, the number of RF chains is 8, which is equal to the number of data streams.
    Fig. \ref{SE_vs_SNR_MU} and Fig. \ref{EE_vs_Ps_MU} show the average bandwidth efficiency and energy efficiency comparison in mmWave MU-MIMO communication systems.
    It should be mentioned that in Fig. \ref{SE_vs_SNR_MU} and Fig. \ref{EE_vs_Ps_MU}, the bandwidth efficiency and energy efficiency are averaged on a per-user basis.
    The SU-SVD curve represents the bandwidth efficiency of the SVD solution for all users.
    Our simulation results show that the twin-resolution phase-shifter network achieves near-optimal average bandwidth efficiency at a significantly reduced power consumption.
    Note that in Fig. \ref{EE_vs_Ps_SU} and Fig. \ref{EE_vs_Ps_MU}, the energy efficiency of the fully digital solution is omitted owing to its extremely high power consumption.

    In Fig. \ref{SE_vs_U_MU}, we investigate the relationship between the bandwidth efficiency and the number of users.
    Each user has $M_s = 2$ data streams.
    The BS is equipped with $N_{\mathrm{RF}} = N_s =  UM_s$ RF chains.
    The SNR in the system is set to 20dB.
    It is observed that the bandwidth efficiency increases as the number of the users increases.
    Furthermore, the bandwidth efficiency of the twin-resolution phase-shifter network is always better than that of the entirely low-resolution phase-shifter network, and it is close to that of the entirely high-resolution phase-shifter network.
    Fig. \ref{EE_vs_U_MU} shows the associated energy efficiency comparison.
    The noise power is $\sigma^2=1$ and the transmit power is $\rho=1$W.
    This demonstrates the superiority of the proposed dynamic twin-resolution phase-shifter network in MU-MIMO systems, when the number of users ranges from 2 to 6.

\subsection{Simulation vs. Theoretical Results}\label{S6.3}
\begin{figure}[t]
\centering
\includegraphics[width=0.45\textwidth]{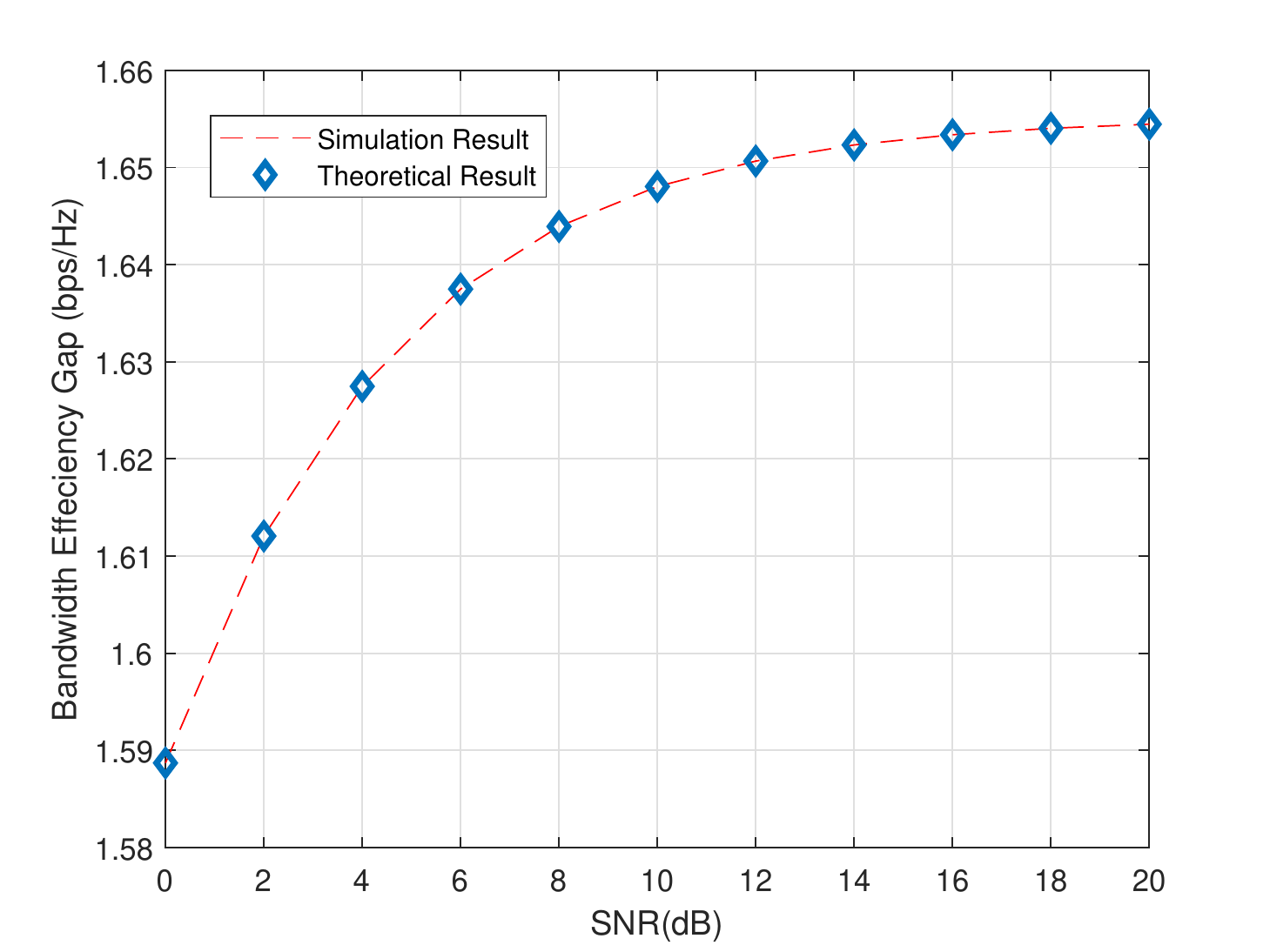}
\caption{Bandwidth efficiency gap vs. SNR in point-to-point mmWave MIMO systems. $N_{\mathrm{BS}}$ = 64, $N_{\mathrm{MS}}$ = 16, $N_{\mathrm{RF}}$ = 4, $N_s$ = 4.}
\label{SEgap_vs_SNR_SU}
\end{figure}

\begin{figure}[t]
\centering
\includegraphics[width=0.45\textwidth]{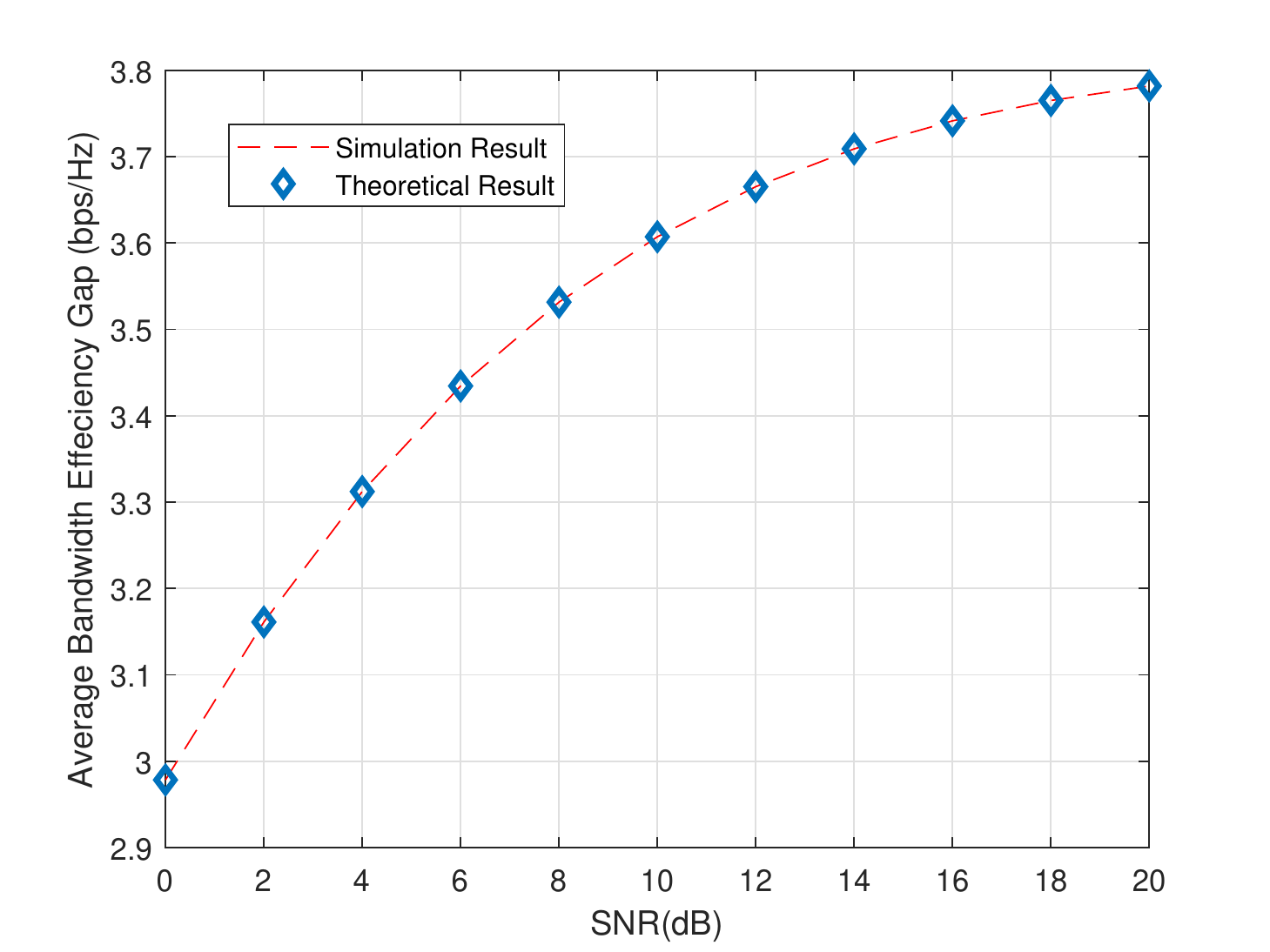}
\caption{Average bandwidth efficiency gap vs. SNR in MU-MIMO systems. $N_{\mathrm{BS}}$ = 64, $N_{\mathrm{MS}}$ = 16, $U$ = 2, $M_s$ = 4, $N_{\mathrm{RF}}$ = 8, $N_s$ = 8.}
\label{SEgap_vs_SNR_MU}
\end{figure}

    Let us now verify the accuracy of our theoretical derivation in Fig. \ref{SEgap_vs_SNR_SU} and Fig. \ref{SEgap_vs_SNR_MU}.
    Two curves are plotted in each figure.
    One of the curves shows the bandwidth efficiency gap acquired by simulations, where the twin-resolution phase-shifter network using 3-bit high-resolution phase shifters and 1-bit low-resolution phase shifters is adopted.
    The other curve represents the bandwidth efficiency gap according to (\ref{R_deltaSVD}) for point-to-point MIMO systems and (\ref{R_deltaSVD_MU}) for MU-MIMO systems.
    It is observed that the simulation results and the theoretical analysis are consistent.
\begin{figure}[t]
\centering
\includegraphics[width=0.45\textwidth]{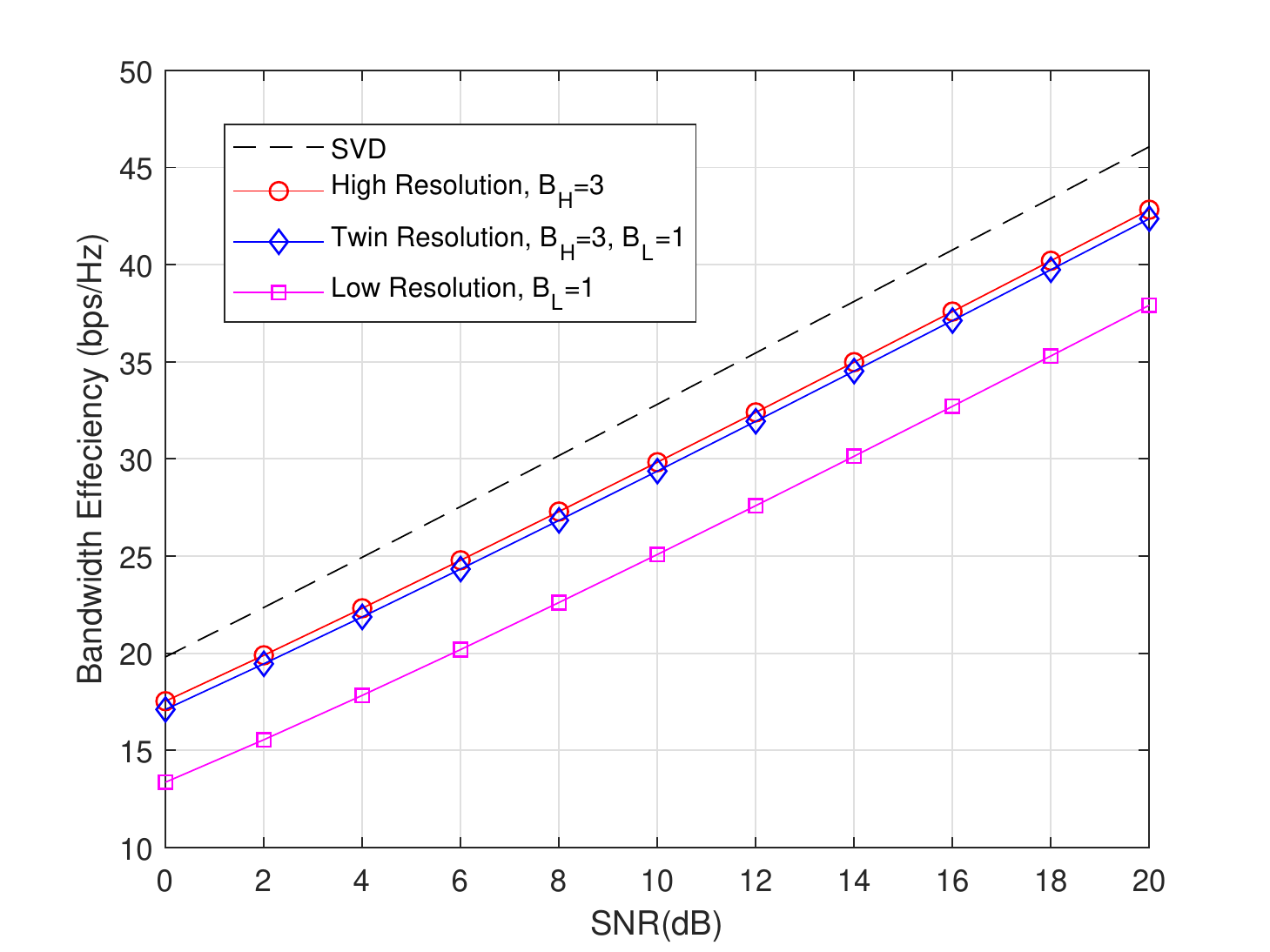}
\caption{{Bandwidth efficiency vs. SNR in wideband mmWave MIMO systems. $N_{\rm BS}$ = 64, $N_{\rm MS}$ = 16, $N_{\rm RF}$ = 4, $N_s$ = 4, $P$ = 128.}}
\label{test_SE_SNRWB}
\end{figure}
\begin{figure}[t]
\centering
\includegraphics[width=0.45\textwidth]{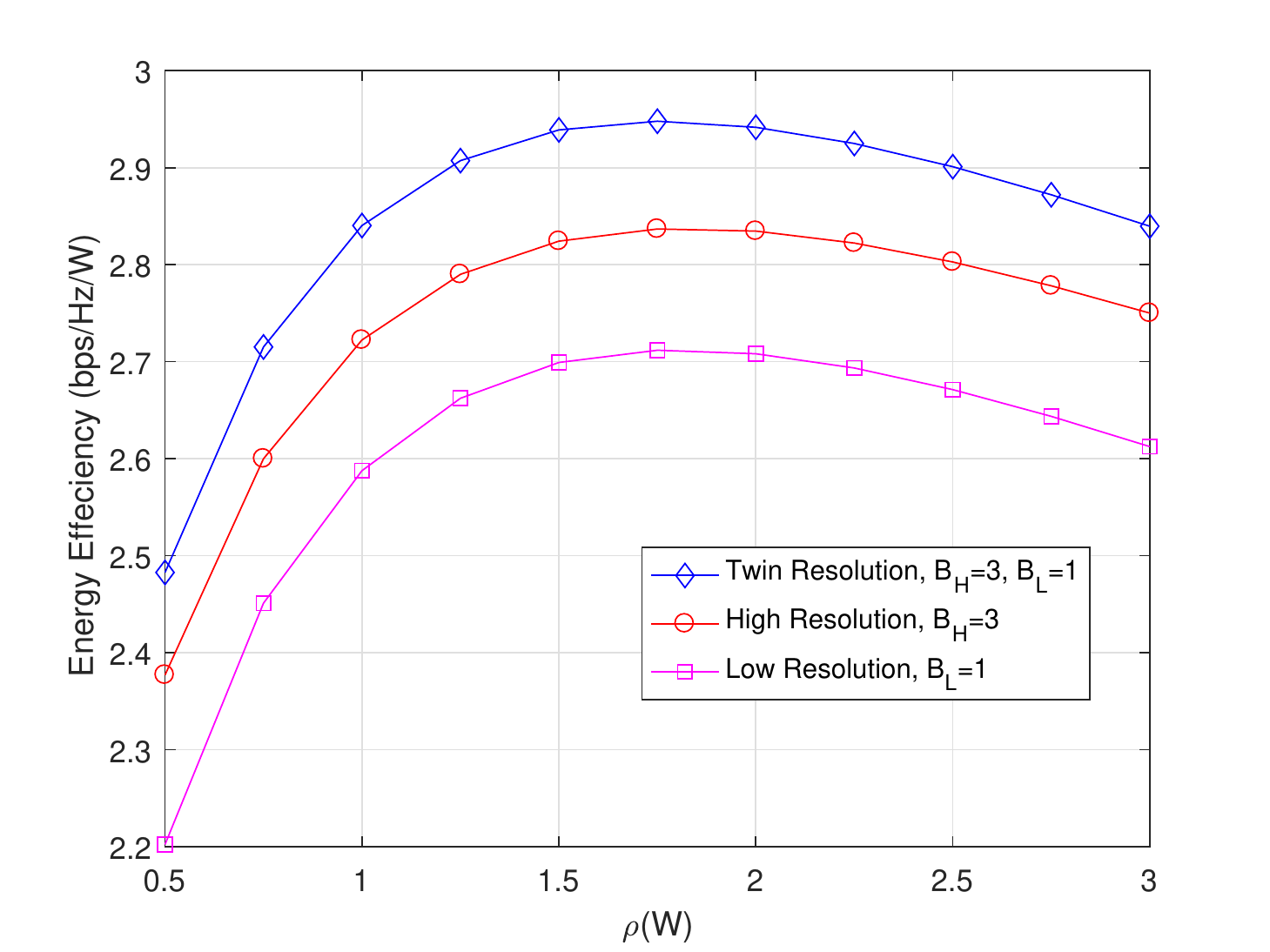}
\caption{{Energy efficiency vs. $\rho$ in wideband mmWave MIMO systems. $N_{\mathrm{BS}}$ = 64, $N_{\mathrm{MS}}$ = 16, $N_{\mathrm{RF}}$ = 4, $N_s$ = 4, $\sigma^2$ = 1, $P$ = 128.}}
\label{test_EE_PsWB}
\end{figure}
\subsection{Simulation Results of Wideband Point-to-point MmWave MIMO Systems}\label{S6.4}
    By considering a frequency-selective channel and the classic OFDM waveform, we extend our dynamic hybrid TPC to a wideband scenario.
    Specifically, we design the baseband TPC matrix by using the SVD method at each subcarrier for a given analog TPC matrix.
For analog TPC design, we can formulate the optimization problem of maximizing the bandwidth efficiency of wideband mmWave MIMO systems as
    \begin{subequations}\label{ObjSU_reWBR23}
    \begin{equation}
    \max_{\mathbf{F}_{\mathrm{RF}}} \quad \frac{1}{P} \sum_{p=1}^P\mathrm{log}_2\left(\left| \mathbf{I}_{N_s} + \frac{\rho}{N_s\sigma^2} \widehat{\mathbf{H}}\left[p\right]\mathbf{F}_{\mathrm{RF}} \mathbf{F}_{\mathrm{RF}}^{\text{H}} \widehat{\mathbf{H}}^{\text{H}}\left[p\right] \right|\right)
    \end{equation}
    \begin{equation}
    s.t. \quad \ \left|\mathbf{F}_{\mathrm{RF}}\left(i,j \right)\right| = \frac{1}{\sqrt{N_\mathrm{BS}}},
    \end{equation}
    \end{subequations}
    where $\widehat{\mathbf{H}}\left[p\right]=\mathbf{W}_{\mathrm{opt}}^{\text{H}}\left[p\right]\mathbf{H}\left[p\right]$ associated with $\mathbf{H}\left[p\right]$ representing the wideband mmWave channel and $ \mathbf{W}_{\mathrm{opt}}\left[p\right] $ the full-digital combiner at the $p$-th subcarrier.
    Then we derive the upper bound for the OF of the problem (\ref{ObjSU_reWBR23}) in the form of
    \begin{align}\label{R_jenson}
    &\frac{1}{P}\sum_{p=1}^P\mathrm{log}_2\left(\left| \mathbf{I}_{N_s} + \frac{\rho}{N_s\sigma^2} \widehat{\mathbf{H}}\left[p\right]\mathbf{F}_{\mathrm{RF}} \mathbf{F}_{\mathrm{RF}}^{\text{H}} \widehat{\mathbf{H}}^{\text{H}}\left[p\right] \right|\right) \nonumber \\
    \overset{\left( a \right)}{\leq}& \mathrm{log}_2 \left(\left| \mathbf{I}_{N_s} + \frac{\rho}{N_s\sigma^2} \mathbf{F}_{\mathrm{RF}}^{\text{H}} \widehat{\mathbf{R}}_{\rm WB} \mathbf{F}_{\mathrm{RF}}\right|\right),
	\end{align}
where $\widehat{\mathbf{R}}_{\rm WB} = \frac{1}{P}\sum_{p=1}^P\widehat{\mathbf{H}}^{\text{H}}\left[p\right]\widehat{\mathbf{H}}\left[p\right]$, and $\left( a \right)$ holds due to Jensen's inequality.
We propose to maximize the bandwidth efficiency upper bound in wideband mmWave MIMO systems \cite{Shen}.
Furthermore, we decompose $\widehat{\mathbf{R}}_{\rm WB}$ as $\widehat{\mathbf{R}}_{\rm WB} = \widehat{\mathbf{R}}_{\rm WB}^{\frac{1}{2}} \left( \widehat{\mathbf{R}}_{\rm WB}^{\frac{1}{2}} \right)^{\text{H}}$ and adopt some basic mathematical transformations.
We can formulate the analog precoder design problem of wideband mmWave MIMO system as
    \begin{subequations}\label{ObjSU_reWB2}
    \begin{equation}
    \max_{\mathbf{F}_{\mathrm{RF}}} \quad  \mathrm{log}_2\left(\left| \mathbf{I}_{N_s} + \frac{\rho}{N_s\sigma^2} \left( \widehat{\mathbf{R}}_{\rm WB}^{\frac{1}{2}} \right)^{\text{H}} \mathbf{F}_{\mathrm{RF}} \mathbf{F}_{\mathrm{RF}}^{\text{H}} \widehat{\mathbf{R}}_{\rm WB}^{\frac{1}{2}}  \right|\right)
    \end{equation}
    \begin{equation}
    s.t. \quad \ \left|\mathbf{F}_{\mathrm{RF}}\left(i,j \right)\right| = \frac{1}{\sqrt{N_\mathrm{BS}}},
    \end{equation}
    \end{subequations}
which has the same form as the problem (\ref{ObjSU_re}).
Thus, the proposed dynamic hybrid precoding method can be adopted for solving (\ref{ObjSU_reWB2}).

    We then embark on numerically evaluating the proposed method in wideband mmWave MIMO systems.
    The parameters are the same as those of the narrow-band point-to-point mmWave MIMO systems except that the bandwidth is set to $2$GHz, the number of subcarriers is set to $128$, and the cyclic prefix length is set to $32$.
    As shown in Fig. \ref{test_SE_SNRWB} and Fig. \ref{test_EE_PsWB}, in wideband mmWave MIMO systems, the twin-resolution phase-shifter network has a near-optimal bandwidth efficiency associated with dramatically improved energy efficiency.
    The trends are the same as those in narrow band scenarios.
    Since the analog TPC is frequency-independent and the channels in all subcarriers have to be taken into account during the design of $\mathbf{F}_{\rm RF}$, the bandwidth efficiency and energy efficiency are slightly lower than those in narrow band scenarios.

    \section{Conclusions}\label{S7}
    In this paper, we proposed a twin-resolution phase-shifter network.
    Furthermore, we proposed a dynamic hybrid TPC method for jointly designing the phase-shifter array pattern and hybrid TPC matrix.
    The proposed method was then slightly modified, when the phase-shifter array pattern was fixed.
    The proposed method was then further extented to MU-MIMO communication systems.
    Our simulation results show that the network advocated achieves near-optimal bandwidth efficiency at a drastically reduced power consumption.
    We also derived the bandwidth efficiency gap between the fully digital solution and the proposed method.
    In the future, we will focus on the energy efficiency optimization of similar techniques.
    \begin{appendices}
          \section{Derivation of (\ref{decomp_1})}
          To derive (\ref{decomp_1}), we have the following equations.
            \begin{align}\label{AppendixB}
            &\mathrm{log}_2\left(\left|\widehat{\mathbf{H}}\mathbf{F}_{\mathrm{RF}} \mathbf{F}_{\mathrm{RF}}^{\text{H}} \widehat{\mathbf{H}}^{\text{H}} \right|\right) \nonumber\\
             = & \mathrm{log}_2\left(\left|\widehat{\mathbf{H}}
            \begin{bmatrix}
            \overline{\mathbf{F}}_{\mathrm{RF}}^{j} & \mathbf{f}_{\mathrm{RF}}^{j}
            \end{bmatrix}
            \begin{bmatrix}
            \left(\overline{\mathbf{F}}_{\mathrm{RF}}^{j}\right)^{\text{H}} \\
            \left(\mathbf{f}_{\mathrm{RF}}^{j}\right)^{\text{H}}
            \end{bmatrix}
            \widehat{\mathbf{H}}^{\text{H}} \right|\right) \nonumber \\
            = & \mathrm{log}_2\left(\left| \mathbf{C}_j + \widehat{\mathbf{H}} \mathbf{f}_{\mathrm{RF}}^{j} \left( \mathbf{f}_{\mathrm{RF}}^{j} \right)^{\text{H}} \widehat{\mathbf{H}}^{\text{H}} \right| \right) \nonumber \\
            \approx & \mathrm{log}_2\left(\left| \mathbf{C}_j \left( \mathbf{I}_{N_s} + \left( \xi\mathbf{I}_{Ns} + \mathbf{C}_j \right)^{-1}\widehat{\mathbf{H}} \mathbf{f}_{\mathrm{RF}}^{j} \left( \mathbf{f}_{\mathrm{RF}}^{j} \right)^{\text{H}} \widehat{\mathbf{H}}^{\text{H}} \right) \right| \right) \nonumber \\
            = & \mathrm{log}_2\left(\left| \mathbf{C}_j \right|\right) \nonumber \\
            \quad  & + \mathrm{log}_2\left(\left| \left( \mathbf{I}_{N_s} + \left( \xi\mathbf{I}_{Ns} + \mathbf{C}_j \right)^{-1}\widehat{\mathbf{H}} \mathbf{f}_{\mathrm{RF}}^{j} \left( \mathbf{f}_{\mathrm{RF}}^{j} \right)^{\text{H}} \widehat{\mathbf{H}}^{\text{H}} \right) \right| \right)\nonumber \\
            \overset{\left( a \right)}{=} & \mathrm{log}_2\left(\left| \mathbf{C}_j \right| \right) + \mathrm{log}_2\left(\left| 1 + \left( \mathbf{f}_{\mathrm{RF}}^{j} \right)^{\text{H}} \mathbf{G}_j \mathbf{f}_{\mathrm{RF}}^{j} \right| \right),
        	\end{align}
        where $\left(a\right)$ holds due to $\mathrm{log}_2\left( \left|\mathbf{I} + \mathbf{X} \mathbf{Y}\right| \right) =\mathrm{log}_2\left(\left| \mathbf{I} +  \mathbf{Y}\mathbf{X} \right| \right)$.
      \end{appendices}
	\bibliographystyle{IEEEtran}
	\bibliography{IEEEabrv,Refference}

\begin{IEEEbiography}[{\includegraphics[width=1in,height=1.25in,clip,keepaspectratio]{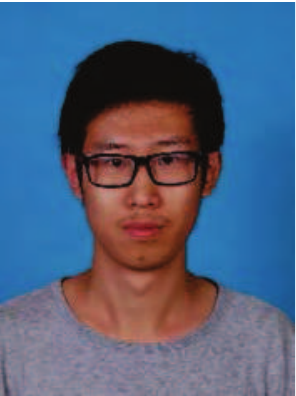}}]
		{Chenghao Feng} received the B.E. degree from the Beijing Institute of Technology, Beijing, China, in 2017, where he is currently pursuing the Ph.D. degree with the School of Information and Electronics. His current research interests include massive MIMO, mmWave/THz communications, energy-efficient communications, intelligent reflecting surface and networks.
\end{IEEEbiography}

\begin{IEEEbiography}[{\includegraphics[width=1in,height=1.25in,clip,keepaspectratio]{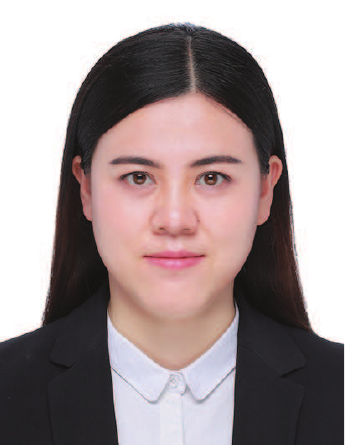}}]
		{Wenqian Shen} received the B.S. degree from Xi'an Jiaotong University, Shaanxi, China in 2013 and the Ph.D. degree from Tsinghua University, Beijing, China. She is currently an associate professor with the School of Information and Electronics, Beijing Institute of Technology, Beijing, China. Her research interests include massive MIMO and mmWave/THz communications. She has published several journal and conference papers in IEEE Transaction on Signal Processing, IEEE Transaction on Communications, IEEE Transaction on Vehicular Technology, IEEE ICC, etc. She has won the IEEE Best Paper Award at the IEEE ICC 2017.
\end{IEEEbiography}

\begin{IEEEbiography}[{\includegraphics[width=1in,height=1.25in,clip,keepaspectratio]{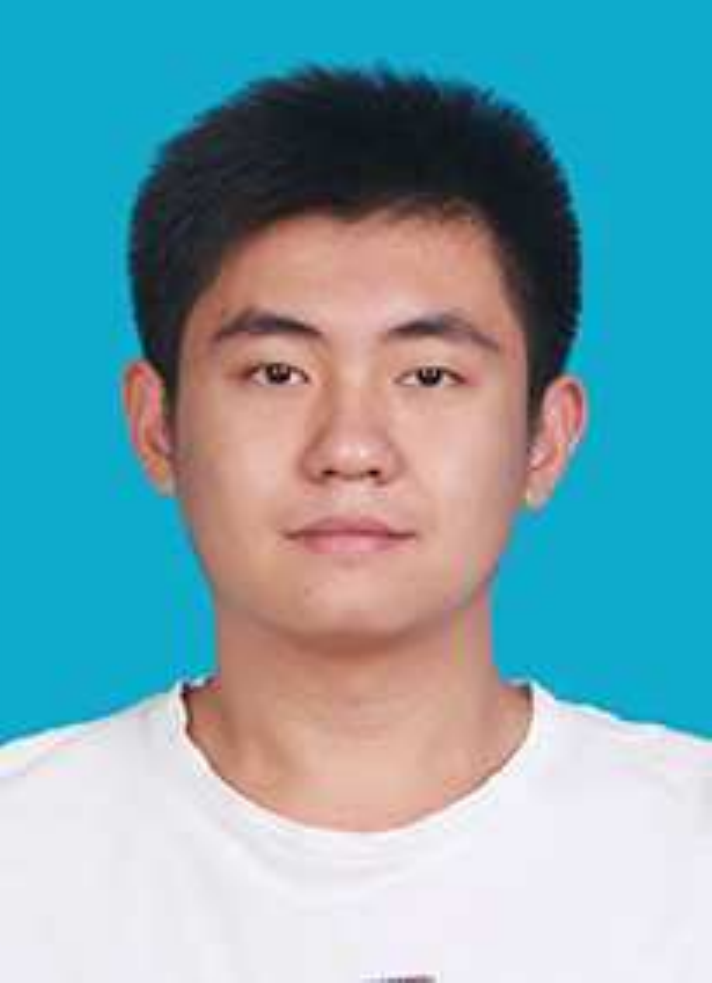}}]
    	{Xinyu Gao}(S'14) received the B.E. degree of Communication	Engineering from Harbin Institute of Technology, Heilongjiang, China in 2014 and the PhD degree of Electronic Engineering from Tsinghua University, Beijing, China in 2019 (with the highest honor). He is currently working as a senior engineer for Huawei Technology, Beijing, China. His research interests include massive MIMO and mmWave communications, with the emphasis on signal processing. He has published more than 20 IEEE journal and conference papers, such as IEEE Journal on Selected Areas in Communications, IEEE Transaction on Signal Processing, IEEE ICC, IEEE GLOBECOM, etc. He has won the WCSP Best Paper Award and the IEEE ICC Best Paper Award in 2016 and 2018, respectively.
\end{IEEEbiography}

	\begin{IEEEbiography}[{\includegraphics[width=1in,height=1.25in,clip,keepaspectratio]{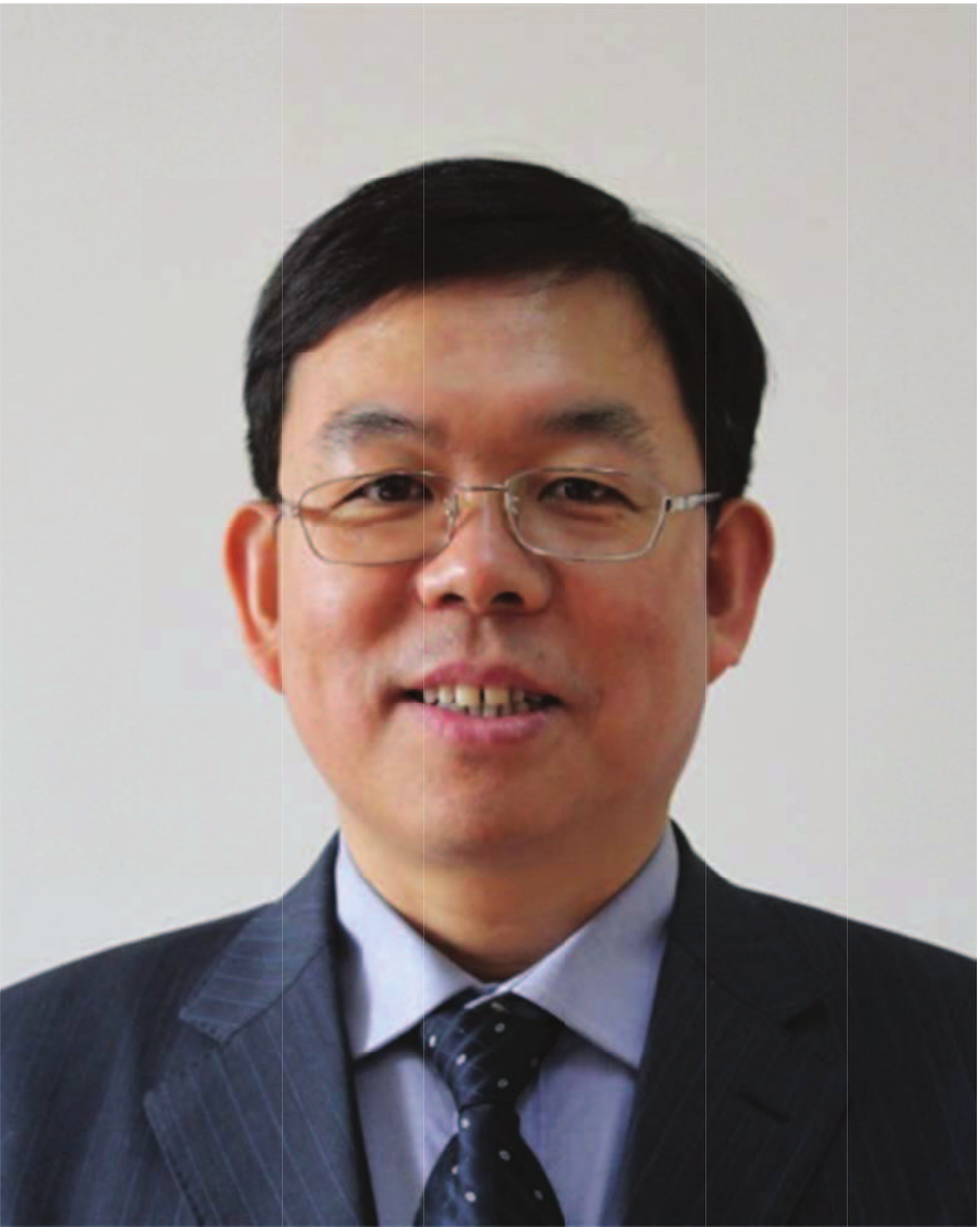}}]
		{Jianping An} (M'08) received the B.E. degree from Information Engineering University in 1987, and the M.S. and Ph.D. degrees from Beijing Institute of Technology, in 1992 and 1996, respectively. Since 1996, he has been with the School of Information and Electronics, Beijing Institute of Technology, where he now holds the post of Full Professor. From 2010 to 2011, he was a Visiting Professor at University of California, San Diego. He has published more than 150 journal and conference articles and holds (or co-holds) more than 50 patents. He has received various awards for his academic achievements and the resultant industrial influences, including the National Award for Scientific and Technological Progress of China (1997) and the Excellent Young Teacher Award by the China's Ministry of Education (2000). Since 2010, he has been serving as a Chief Reviewing Expert for the Information Technology Division, National Scientific Foundation of China. Prof. An's current research interest is focused on digital signal processing theory and algorithms for communication systems.
	\end{IEEEbiography}

\begin{IEEEbiography}[{\includegraphics[width=1in,height=1.25in,clip,keepaspectratio]{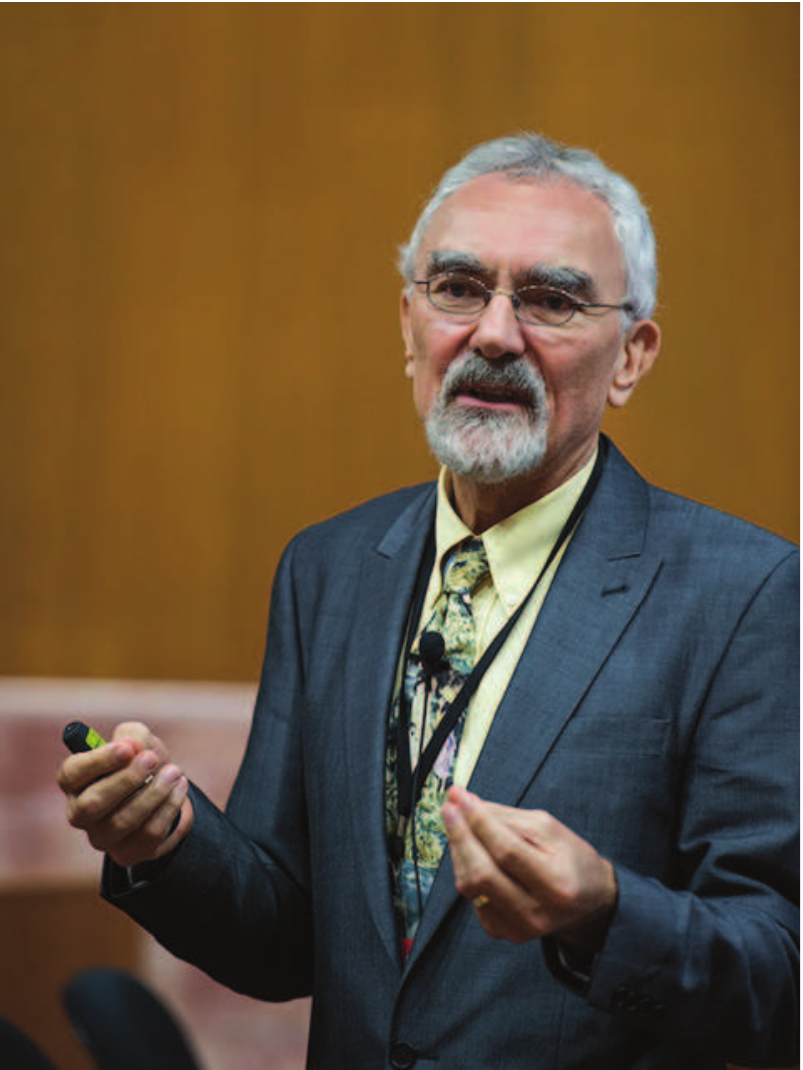}}]
		{Lajos Hanzo} (http:\/\/www-mobile.ecs.soton.ac.uk,
https:\/\/en.wikipedia.org\/wiki\/Lajos\_Hanzo) (FIEEE'04, Fellow of the
Royal Academy of Engineering F(REng), of the IET and of EURASIP),
received his Master degree and Doctorate in 1976 and 1983,
respectively from the Technical University (TU) of Budapest. He was
also awarded the Doctor of Sciences (DSc) degree by the University of
Southampton (2004) and Honorary Doctorates by the TU of Budapest
(2009) and by the University of Edinburgh (2015).  He is a Foreign
Member of the Hungarian Academy of Sciences and a former
Editor-in-Chief of the IEEE Press.  He has served several terms as
Governor of both IEEE ComSoc and of VTS.  He has published 1900+
contributions at IEEE Xplore, 19 Wiley-IEEE Press books and has helped
the fast-track career of 123 PhD students. Over 40 of them are
Professors at various stages of their careers in academia and many of
them are leading scientists in the wireless industry.
\end{IEEEbiography}

\end{document}